\documentclass[traditabstract]{aa} 
\usepackage{epsfig}
\usepackage{graphicx}
\usepackage{xcolor}
\usepackage{txfonts}
\usepackage{amssymb}
\usepackage{amsfonts}
\usepackage{natbib}
\usepackage{hyperref}
\hypersetup{
  colorlinks=true, 
  urlcolor=blue, 
  linkcolor=blue, 
  citecolor=blue 
}
\usepackage{multirow}

\bibpunct{(}{)}{;}{a}{}{,}
\newcommand{\be}{\begin{equation}}
\newcommand{\ee}{\end{equation}}
\newcommand{\bea}{\begin{eqnarray}}
\newcommand{\eea}{\end{eqnarray}}

\newcommand{\red}[1]{\textcolor{red}{#1}}

\begin{document}

\title{Application of hydrostatic local thermodynamic equilibrium atmosphere models to interpretations of supersoft X-ray source spectra}
\author{
V.~F. Suleimanov
\and 
A.~S. Tavleev
\and
V. Doroshenko
\and
K. Werner}


\institute{
Institut f\"ur Astronomie und Astrophysik, Kepler Center for Astro and
Particle Physics, Universit\"at T\"ubingen, Sand 1, 72076 T\"ubingen, Germany\\ \email{suleimanov@astro.uni-tuebingen.de}
}

\date{Received xxx / Accepted xxx}

   \authorrunning{Suleimanov et al.}
   \titlerunning{Model spectra of super-soft sources}

\abstract
{Supersoft X-ray sources (SSSs) are accreting white dwarfs (WDs) with stable or recurrent thermonuclear burning on their surfaces. High-resolution X-ray spectra of such objects are rather complex, often consist of several components, and are difficult to interpret accurately. The main emission source is the hot surface of the WD and the emergent radiation can potentially be described by hot WD model atmospheres. We present a new set of such model atmosphere spectra computed in the effective temperature range from $100\rm\,kK$ to $1000\rm\,kK$, for eight values of surface gravity and three different chemical compositions. These compositions correspond to the solar one as well as to the Large and Small Magellanic Clouds, with decreased heavy element abundances, at one-half and one-tenth of the solar value. The presented model grid covers a broad range of physical parameters and, thus, it can be applied to a wide range of objects. It is also publicly available in XSPEC~format. As an illustration, we applied it here for the interpretation of \textit{Chandra}  and XMM grating spectra of two classical SSSs, namely, CAL~$83$ (RX~J0543.5$-$6823) and RX~J0513.9$-$6951. The obtained effective temperatures and surface gravities of $T_{\rm eff} \approx 560$\,kK, $\log g \approx 8.6-8.7$, and $T_{\rm eff} \approx 630\,{\rm kK},   \log g \approx 8.5-8.6$,  respectively, are in a good agreement with previous estimations for both sources. 
The derived WD~mass estimations are within $1.1-1.4\,M_\odot$ for CAL\,83
and $1.15-1.4\,M_\odot$ for RX~J0513.9$-$6951.
The mass of the WD in CAL~$83$ is consistent with the mass predicted from the respective model of recurrent thermonuclear burning. }

\keywords{accretion, accretion discs -- stars: white dwarfs  --  stars: atmospheres -- methods: numerical  -- X-rays: binaries -- X-ray: individuals: CAL~$83$, RX~J0513.9$-$6951}

\maketitle
%

\section{Introduction}
Supersoft sources~(SSSs) are close binary systems that involve a white dwarf~(WD) undergoing accretion
as the primary component. The accretion rate is high enough for (quasi-)stable thermonuclear burning to occur on the WD~surface \citep{1992A&A...262...97V}. These sources were initially discovered in the Large Magellanic Cloud (LMC) by the \textit{Einstein} observatory in the early 1980s~\citep{1981ApJ...248..925L} and were classified as a distinct class of sources following ROSAT observations~\citep{1991Natur.349..579T,1991A&A...246L..17G}. 
The X-ray spectra of SSSs observed by ROSAT were very soft, with blackbody temperatures $kT\sim30-40\rm\,eV$. Formally, these temperatures and the observed fluxes led to luminosities exceeding the Eddington limit for a solar mass object at the LMC distance. This issue was resolved through the implementation of hot WD~model atmosphere spectra for interpreting the X-ray spectra of SSSs~\citep{1994A&A...288L..45H}. The results of early investigations of SSSs in this context were presented and discussed by~\citet{1997ARA&A..35...69K}.

In addition to the classical SSSs, super-soft X-ray phases occur during nova explosions. The hot WD with ongoing thermonuclear burning on the surface becomes visible in soft X-rays after the dispersal of an optically thick envelope \citep[see, e.g.,][and references therein]{1999A&A...347L..43K, 2001A&A...373..542O, 2003ApJ...594L.127N, 2011ApJS..197...31S}.

Model atmospheres and their theoretical spectra are important ingredients in investigating the nature of SSSs and determining their basic parameters. The first models were computed \citep{1994A&A...288L..45H} under the assumption of local thermodynamic equilibrium (LTE), but non-LTE model atmospheres of hot WDs were subsequently developed as well \citep{1997A&A...322..591H}. 
The first non-LTE models did not include spectral lines, but later a number of spectral lines were incorporated~\citep{1999A&A...346..125H}, using the publicly available code \texttt{TLUSTY}~\citep{1988CoPhC..52..103H}. 
For the interpretation of the X-ray spectrum of the bright classical SSS CAL~$83$, \citet{2005ApJ...619..517L} used more accurate non-LTE model atmospheres. 
Such models were further developed to analyse the super-soft phase of nova V$4743$~Sgr~\citep{2010ApJ...717..363R}, where various chemical compositions of the atmospheres as well as the metal-line blanketing effects were considered. These models were calculated using the T\"ubingen non-LTE Model Atmosphere Package~\citep[\texttt{TMAP}, ][]{2003ASPC..288...31W, 2010AN....331..146R}. Unfortunately, the accurate computation of non-LTE models is very computation-time expensive, so only small regions of physical parameter space can be probed and models are typically tailored to individual objects.

{In addition to non-LTE effects in hydrostatic model atmospheres, expansion due to radiation pressure force in spectral lines could also be important for high temperatures reached in the SSSs. 
Expanding model atmospheres were thus considered and computed by ~\cite{2010AN....331..175V} and \cite{2012ApJ...756...43V} using the publicly available code \texttt{PHOENIX}~\citep[see, e.g.,][]{1997ApJ...483..390H}. Further observational evidence of SSS~atmosphere expansion was presented by~\citet{2010AN....331..179N}. Again, modelling these effects, especially in a non-LTE approximation, is challenging, precluding a full exploration of the parameter space.
 
 On the other hand, LTE model atmospheres of hot WDs can be computed much faster, and, despite their simplicity, were successfully used to interpret ROSAT spectra of SSSs \citep{2003ARep...47..186I, 2003ARep...47..197S}, and a few SSSs found in M81 \citep{2002ApJ...574..382S}.
However, the high-resolution grating spectra obtained by \textit{Chandra} and \textit{XMM-Newton} observatories demonstrated that the hot WD model spectra cannot completely describe observations and reproduce only the common spectral shape with separate strong absorption lines in some cases. This holds true for both hydrostatic and expanding non-LTE model atmospheres~\citep{2020AdSpR..66.1202N}, suggesting that there may be some missing physical processes and additional X-ray emission sources, such as inner accretion disc or boundary layers. 
Nevertheless, basic SSS~parameters appear to be well recovered. Moreover, LTE atmospheres have some advantages compared to sophisticated non-LTE expanding model atmospheres. Hydrostatic LTE model atmospheres are simpler to compute and, what is more important, allow us to consider an almost unlimited number of chemical elements, ions, and spectral lines. Therefore, extended sets of LTE~model atmospheres computed for various chemical compositions could potentially be useful for the approximate estimation of basic physical parameters of hot~WDs in~SSSs and the chemical composition of their atmospheres. The latter problem is especially important, for instance, for the super-soft phases of nova explosions.

Here, we present a set of hot WD~model spectra computed for three chemical compositions of the atmospheres: the solar abundance, typical for the Galaxy; the solar H/He mix with the heavy element abundances reduced by a factor of 2 \citep[typical for the LMC, see, e.g.,][]{2002A&A...396...53R}; and by a factor of 10 \citep[typical for the Small Magellanic Cloud (SMC), see, e.g.,][]{2008AJ....136.1039C} in comparison with the solar abundance. We employed the approach previously used for computing the boundary layer spectrum in the dwarf nova SS~Cyg~\citep{2014A&A...571A..55S}. A preliminary version of this set of models has also already been used to fit a nova spectrum in fireball phase~\citep{2022Natur.605..248K} 
 
The remainder of this paper is organised as follows. In Sect.~\ref{sec:method} we describe the method of computation of the atmospheric models, in particular, our new approach to calculate photoionisation opacities. In Sect.~\ref{sec:results} we first present the characteristics of the resulting new set of model atmosphere spectra. Then we apply the obtained set to analyse X-ray spectra of the classical~SSSs CAL~$83$ and RX~J0513.9$-$6951. The results of this analysis are discussed in the context of WD evolution in Sect.~\ref{sec:discussion}. We summarize our results in Sect.~\ref{sec:summary}.

\section{Method}
\label{sec:method}

\subsection{Basic assumptions}

In this work, we use a standard method for computing hydrostatic plane-parallel model atmospheres \citep[see, e.g.,][]{1978stat.book.....M} and the code based on the popular Kurucz's code \texttt{ATLAS}~\citep{1970SAOSR.309.....K, 1993yCat.6039....0K}, modified for high temperatures by~\citet{2003ARep...47..186I}; see also~\citet{2013ASPC..469..349S, 2014A&A...571A..55S}.
The general temperature correction scheme for model computation is the same as in \texttt{ATLAS} and the main changes concern the calculation of the opacities and number densities .

We took into account the 15~most abundant chemical elements and the number densities of all ionization and excitation states of all the ions were computed, using the Boltzmann and Saha equations, assuming LTE. We considered pressure ionization and level dissolution effects using the occupation probability formalism~\citep{1988ApJ...331..794H}, as described by~\citet{1994A&A...282..151H}. We used the spectral line list together with necessary physical parameters such as $gf$ values and the energies of the low-energy levels presented in the CHIANTI database \citep{1997A&AS..125..149D, 2021ApJ...909...38D}. The shapes of line absorption opacity are considered as Voigt profiles. Classical damping broadening was used together with the T\"ubingen approximation for Stark broadening \citep{1971Obs....91..139C, 2003ASPC..288...31W}. The lines of hydrogen-like ions are considered using Griem's theory of linear Stark broadening \citep{1960ApJ...132..883G, 1967ApJ...148..547G}. The modified Kurucz's subroutine \citep{1970SAOSR.309.....K} was used for this purpose. 
We note, however, that the Stark broadening for the lines of the highly charged hydrogen-like ions is small ($\sim Z^{-5}$) in comparison with the radiation damping.
A low microturbulent velocity value of $2\,{\rm km\,s}^{-1}$ was added to the thermal velocity of ions to account for Doppler line broadening\footnote{A local subset with the microturbulent velocity equal to the local sound speed was computed for the CAL\,83 spectrum fitting (see below). The fitting result changed insignificantly.}. 

Formally, the radiation pressure force in spectral lines, $g_{\rm rad}$, exceeds the gravity, $g,$ at the upper atmosphere layers and wind model atmospheres have to be used. However, we employed a simple trick, suggested by~\citet{2003ARep...47..186I}, to keep the atmosphere in hydrostatic equilibrium. It was assumed that the gas pressure equals $10\%$~of the total pressure, $P_{\rm g} = 0.1 P_{\rm tot} = 0.1 gm$, at all depths where $g_{\rm rad} > g$. Here, $m$ is a column density ($dm = -\rho dz$), the independent depth variable in our model, $\rho$ is the plasma density, and $z$ is the geometrical depth. This assumption, in fact, corresponds to a specific velocity law in the upper atmosphere layers, $\rm{v}(z) \sim \rho(z)^{-1}$. This approach is the simplest way to avoid the hydrostatic equilibrium violation. All other reasonable methods significantly complicate the atmosphere modeling. For instance, this would require consideration of spherical moving atmospheres, including radiation transfer. That is why we limited ourselves to the proposed method.

To calculate the bound-free opacities from the atomic ground states of all ions, we utilized the procedure presented by~\citet{1996ApJ...465..487V}. The most significant changes for our present model calculations are associated with the photoionization opacities from the excited energy levels of heavy element ions. The method for estimating these opacities and the obtained results are presented in the next subsection.

The free-free opacities of all ions are calculated under the assumption that the ion's electric field is a Coulomb field of charge, $Ze$, corresponding to the ionic charge number, $Z$, with $e$ being the elementary charge. The corresponding Gaunt factors are computed following \citet{1998MNRAS.300..321S}.
 
\subsection{Photoionization from excited energy levels}
Two different approaches for the photoionization and free-free opacities from the excited levels can be used to consider excited levels of hydrogen-like and helium-like ions and of other ions separately. The approach suggested by~\citet{1961ApJS....6..167K} and the corresponding Kurucz subroutine~\citep{1993KurCD..13.....K} was used for these ions. The same approach was applied to carbon model atmospheres of neutron stars~\citep{2014ApJS..210...13S}. 
Photoionization cross-sections from excited levels were computed assuming a Coulomb field of a point-like charge. This approach provides excellent results for hydrogen-like ions, as shown in Fig.~\ref{figNeX}. However, it is not strictly correct for helium-like ions, as the inner (non-excited) electron is included in the point-like effective charge. This leads to some deviations from the exact cross-sections (see Fig.~\ref{fig:Helike}). Nevertheless, the obtained accuracy is acceptable for our purposes. We included five excited levels for hydrogen-like ions and ten levels for helium-like ions. 

\begin{figure}
\centering
\includegraphics[angle=0,scale=1.0]{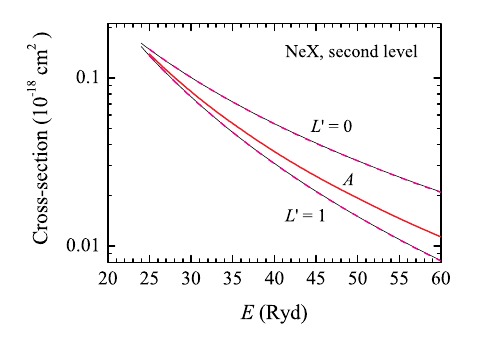}
\caption{Comparison of the photoionization cross-sections of the lowest excited levels of the hydrogen-like ion~\ion{Ne}{X} with the orbital quantum numbers $L'=0$~and~$1$, computed by~OP (thin black curves) and by the used approximation (dashed magenta curves). The $g$-factor weighted averaged over both sub-level cross-sections is shown by the red curve. }
\label{figNeX}
\end{figure}

\begin{figure}
\centering
\includegraphics[angle=0,scale=1.0]{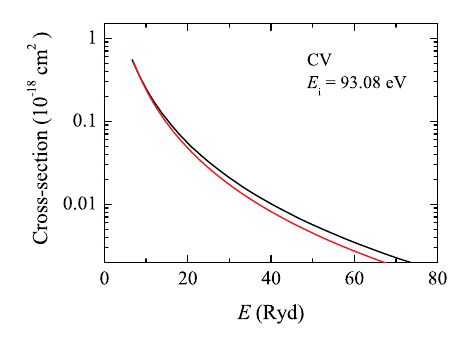}
\includegraphics[angle=0,scale=1.0]{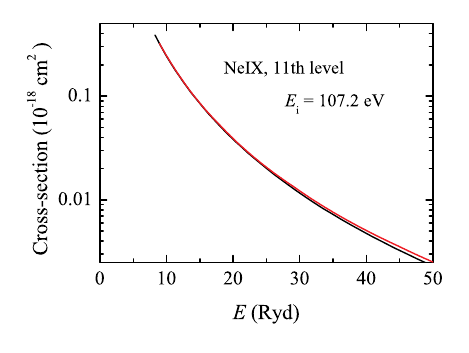}
\caption{Comparison of the photoionization cross-sections from the first excited level of the helium-like ion~\ion{C}{V} \textit{(top panel)} and~11th excited level of helium-like ion~\ion{Ne}{IX} \textit{(bottom panel)}, computed by~OP (solid black curves) and by the used approximation (red curves). The excitation energies of the levels are also shown.}
\label{fig:Helike}
\end{figure}

Another approach is used for ions with more than two electrons. The absorption opacities from the excited levels of these ions were computed by the Opacity Project~\citep[OP,][]{1994MNRAS.266..805S} and are available as numerical tables in the TOPBase database\footnote{\url{https://cdsweb.u-strasbg.fr/topbase/topbase.html}}. We used these tables for creating simplified analytical fits and implemented these fits in our code. Only a few levels with the lowest excitation energies were considered, typically not exceeding a quarter of the ionization energy. Although it should be noted that the choice of the levels under consideration was quite arbitrary. 

The cross-section for each considered level was fitted using the expression suggested by~\citet{1970SAOSR.309.....K}:
\be \label{eq:sig}
  \sigma(E) = \sigma_{\rm th} \left(A \bar E ^{p/2}+(1-A)\bar E^{1+p/2}\right),
\ee
where $\bar E =E_{\rm th}/E$, $E$ is the photon energy, $E_{\rm th}$ is the threshold photoionization energy from the given level, and $\sigma_{\rm th}$ is the photoionization cross-section at the photoionization threshold. The cross-section 
is a fitting parameter as well as the parameters $A$ and $p$. 
The values $E_{\rm th}$ and the obtained fitting parameters  $\sigma_{\rm th}$, $A$ and~$p$ together with the level statistical weight~$g$ for the approximation~(\ref{eq:sig}) for lower excited levels of C, N, O, Ne, Na, Mg, Al, Si, S, Ar, Ca, and Fe ions are publicly available\footnote{\url{https://doi.org/10.5281/zenodo.10277303}}.
Some examples of the fitting are shown in Figs.~\ref{figCII}~and~\ref{figCIV}. We note that our smooth fits ignore the auto-ionization resonances (see Fig.~\ref{figCII}) and OP~cross-sections significantly differ from the old ones used in~\texttt{ATLAS9}. Simpler ions with one electron in the outer shell have smooth cross-section dependencies, approximated with a high level of accuracy (Fig.~\ref{figCIV}).   

The excited levels of the helium-like iron ion have more complicated cross-section dependencies on the photon energy, consisting of two parts. The low-energy one just above the photoionization threshold is well fitted with the approach used for other helium-like ions. However, at some energy, approximately ten times larger than the photoionization threshold energy, the cross-section sharply increases. We considered this second part as an additional photoionization edge and approximated it using Eq.~(\ref{eq:sig}). The comparison of these double fitting with the computed cross-section for one of the excited levels of~\ion{Fe}{XXV} is shown in Fig.~\ref{figFeXXV}. 

\begin{figure}
\centering
\includegraphics[angle=0,scale=1.0]{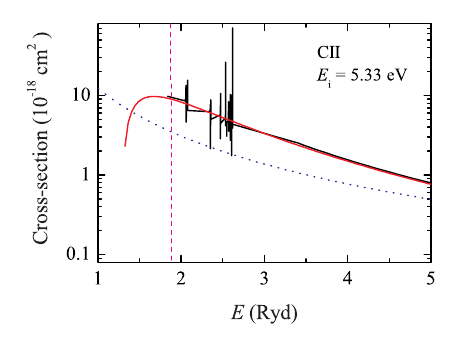}
\includegraphics[angle=0,scale=1.0]{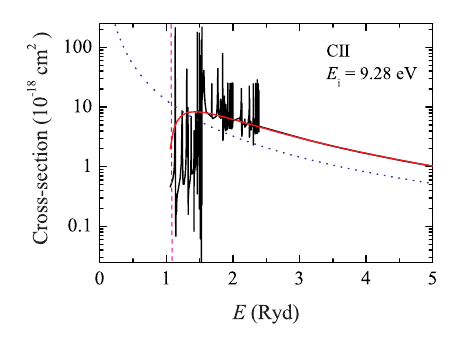}
\caption{Comparison of the photoionization cross-sections from two excited levels of \ion{C}{II}.  Cross-sections computed by~OP (solid black curves), the approximation used in \texttt{ATLAS9} (dashed blue curves), and our fits (red curves). Vertical dashed lines correspond to the ionization thresholds.}
\label{figCII}
\end{figure}

\begin{figure}
\centering
\includegraphics[angle=0,scale=1.0]{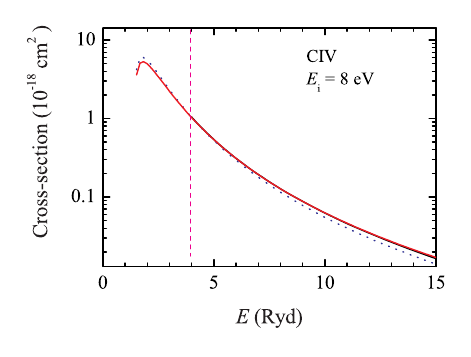}
\caption{Same as in Fig.~\ref{figCII} for the first excited level of \ion{C}{IV}. Black and red curves overlap and are close also to \texttt{ATLAS9} approximation (blue dashed curve) here.}
\label{figCIV}
\end{figure}

\begin{figure}
\centering
\includegraphics[angle=0,scale=1.0]{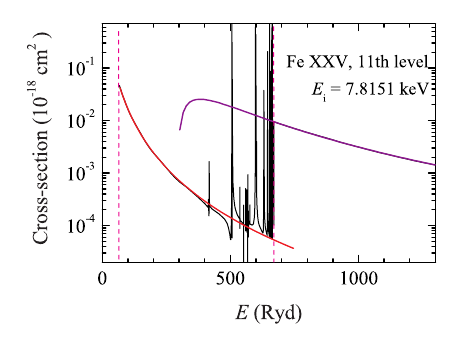}
\caption{Comparison of the photoionization cross-sections from the~$11^{\rm th}$ excited level of the helium-like ion~\ion{Fe}{XXV}. Shown are cross-sections computed by~OP (solid black curve) and the used approximation by two fitting functions (red curves). More details in the text. The fitting function thresholds are indicated by the vertical dashed magenta lines.}
\label{figFeXXV}
\end{figure}

\section{Results}
\label{sec:results}

\subsection{Model atmospheres}

We computed three sets of two-parameter plane-parallel hot WD~model atmospheres with different chemical compositions. All the sets have the same solar hydrogen-helium mix and different abundances of heavy elements: the solar one ($A=1$), half the solar abundance ($A=0.5$), which corresponds to the LMC heavy element abundance, and ten times less than the solar one ($A=0.1$), which is close to the heavy element abundance in the SMC.
The model parameters of each set are the effective temperature, $T_{\rm eff}$ ($100$ to $1000\,\rm kK$ in steps of $25\,\rm kK$), and $\Delta\log g = \log g - \log g_{\rm Edd}$, which characterises the distance of the model from the Eddington limit:
 \be
    \log g_{\rm Edd} = \log (\sigma_{\rm e}\sigma_{\rm SB}\,T_{\rm eff}^4\,c^{-1}) = 4.818 + 4\,\log (T_{\rm eff}/10^5\,{\rm K}),
 \ee
where $\sigma_{\rm SB}$ is the Stefan-Boltzmann constant, $c$ is the speed of light, and $\sigma_{\rm e} = 0.2(1+X) \approx 0.34\,{\rm cm}^2\,{\rm g}^{-1}$. Here, $X\approx 0.7374$ is the hydrogen mass fraction. The parameter $\Delta\log g$ has eight values in the grid: $0.1, 0.2, 0.4, 0.6, 1.0, 1.4, 1.8, {\rm and}\, 2.2$. Altogether, 296 spectra were computed in every set and implemented as table model\footnote{\red{\url{https://github.com/HEASARC/xspec_localmodels/tree/master/sss_atm}}} for~XSPEC\,\footnote{\url{https://heasarc.gsfc.nasa.gov/docs/xanadu/xspec/}}~\citep{Arnaud1996, XSPEC_ascl}.

\begin{figure}
\centering
\includegraphics[width=0.99\columnwidth]{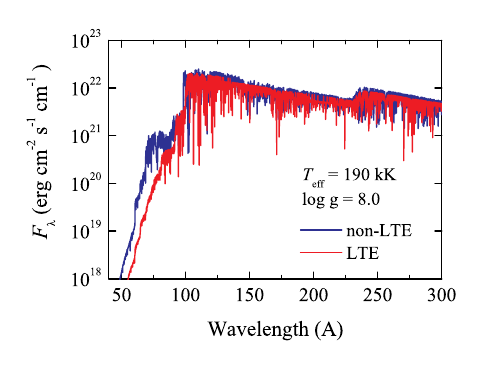}
\caption{Comparison of the emergent spectra of the model atmospheres computed by the code presented here (red curves), and the \texttt{TMAP} non-LTE code~\citep[][blue curves]{2003A&A...403..709R}. All models have solar chemical composition, 
$T_{\rm eff} =190\,{\rm kK},\, \log g =8.0$.}
\label{fig:RS}
\end{figure}

The properties of the model atmosphere spectra are illustrated in Figs.~\ref{fig:RS}--\ref{fig:500gA}. A comparison of the model spectrum computed with the LTE~code described above with the hottest non-LTE model atmosphere spectrum presented by~\citet{2003A&A...403..709R} is shown in Fig.~\ref{fig:RS} (top panel). Both spectra were binned with a $0.1$\AA\ wide step. The general shapes of the spectra are similar, and some differences appear at small wavelengths where the emergent flux is insignificant. It could be due to both non-LTE effects and the difference in the number of excited levels taken into consideration. The lists of the used spectral lines are also different, leading to diverse contributions of the lines to the spectra.

\begin{figure}
\centering
\includegraphics[width=0.99\columnwidth]{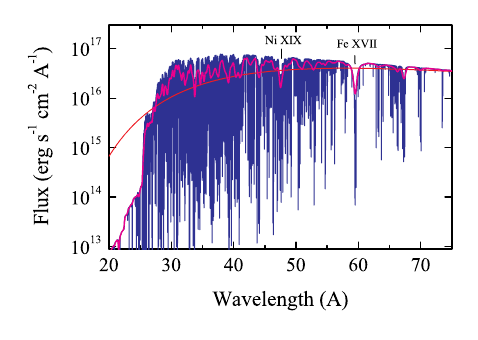}
\includegraphics[width=0.99\columnwidth]{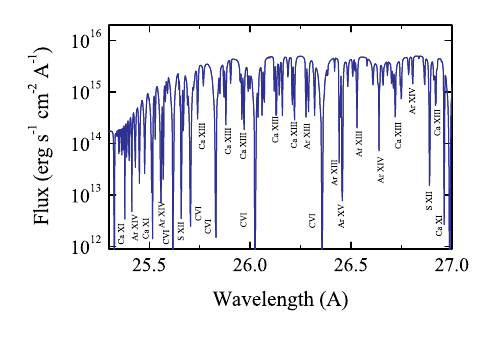}
\caption{ Spectrum of the model atmosphere with $T_{\rm eff}=500\,{\rm kK}, \Delta\log g = 0.6$ and $A=0.5$ (top). The Planck function with the same temperature is also shown by the red curve. 
 The spectrum convolved with a Gaussian kernel with the {\it Chandra} resolution, $R=300,$ is shown in magenta. 
\textit{} Detail of the upper spectrum near the \ion{C}{V} ground state edge with spectral line identifications (bottom).}
\label{fig:500kK}
\end{figure}

The model spectrum of the fiducial model atmosphere ($T_{\rm eff}=500$\,kK, $\Delta\log g =0.6$, and $A=0.5$) is shown in Fig.~\ref{fig:500kK}. The upper panel presents, in the observed soft X-ray wavelength range, the spectrum, the corresponding Planck function, and the spectrum convolved with a Gaussian kernel (the kernel resolution $R=300$ corresponds the \textit{Chandra} grating resolution). The convolved spectrum demonstrates emission line-like structures, which are, in fact, parts of the spectrum with a small number of spectral lines. A detail of the spectrum near the \ion{C}{V} ground state photoionization edge is shown in the bottom panel. The most prominent absorption lines are identified. It is clearly seen that along with the Lyman-like line series of the hydrogen-like \ion{C}{V} ion, there are numerous absorption lines of ions of heavier elements with about $10$~electrons in their shells. We note that the absorption lines of these ions dominate in the overall spectrum.  

\begin{figure}
\centering
\includegraphics[width=0.99\columnwidth]{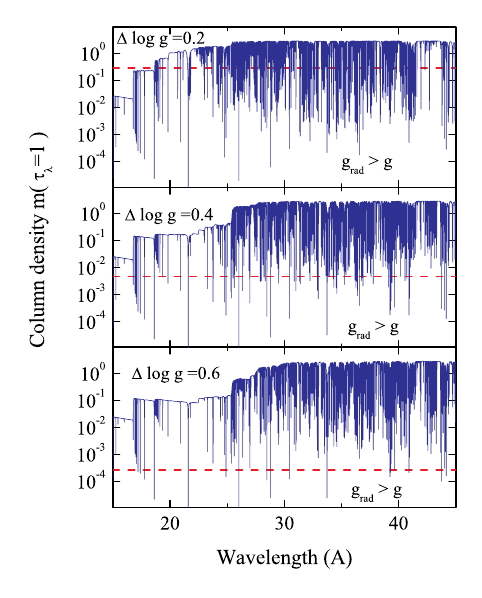}
\caption{Comparison of the emergent spectra formation depth ($m(\tau_\lambda=1)$) with the wind dominating boundary (dashed red lines) for the model atmospheres with $T_{\rm eff}=500$\,kK, $A=0.5$, and three different $\Delta\log g = 0.2$ (top ), $0.4$ (middle) and $0.6$ (bottom panel). Atmospheric layers below the dashed red lines are dominated by the wind. }
\label{fig:grad}
\end{figure}

There is an important note about our assumption that $P_{\rm g} = 0.1 g\,m$ at the upper atmospheric layers where the radiation force due to numerous spectral lines is greater than the WD~gravity ($g_{\rm rad} > g$). In fact, most of the escaping radiation aside from the strong line cores, forms in the hydrostatic layers, at least for models with $\Delta \log g \ge 0.4$, see Fig.~\ref{fig:grad}. The depths where the escaping radiation forms, namely $m(\tau_\lambda = 1)$, are shown there. Here, $\tau_\lambda$~is the optical depth at a wavelength~$\lambda$. The boundaries between the hydrostatic and the wind-dominated layers ($g_{\rm rad}=g$) are also shown. In the model with $\Delta \log g = 0.2,$ a significant part of the lines forms in the wind layers and it is even more prominent in the models with less $\Delta \log g$. Therefore, we conclude that the used assumption is acceptable for models with $\Delta \log g \ge 0.4$. The models with smaller $\Delta \log g$ should be used with caution.

\begin{figure}
\centering
\includegraphics[width=0.99\columnwidth]{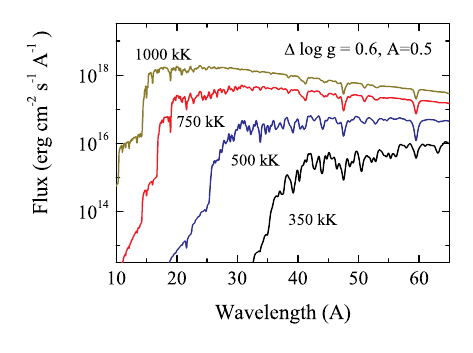}
\caption{Model spectra with various $T_{\rm eff}$ values, fixed $\Delta \log g=0.6,$ and $A=0.5$.}
\label{fig:spT}
\end{figure}

The dependence of the emergent spectrum on the effective temperature with fixed $\Delta \log g = 0.6$ and $A=0.5$ is shown in Fig.~\ref{fig:spT} (here and in the next figure, the convolved spectra are shown). As expected, with increasing $T_{\rm eff}$ the spectrum becomes harder. Decreasing the surface gravity also makes the emergent spectrum harder if other parameters are not changed (Fig.~\ref{fig:500gA}, top panel). The influence of the chemical composition is not significant, although the emergent spectrum becomes slightly harder at lower~$A$~(Fig.~\ref{fig:500gA}, bottom panel).

\begin{figure}
\centering
\includegraphics[width=0.99\columnwidth]{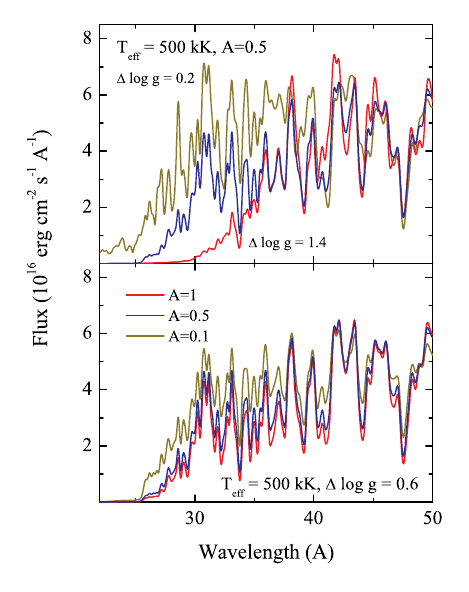}
\caption{Comparison of the fiducial model spectra ($T_{\rm eff} =500\,{\rm kK}, \Delta\log g=0.6,\, A=0.5$) with the spectra computed for different $\Delta\log g$ (top) and~$A$~(bottom panel).}
\label{fig:500gA}
\end{figure}

\subsection{Application of the model to supersoft sources}

The resulting model atmosphere grids cover large surface gravity and temperature ranges and can be used to estimate physical parameters of hot WDs, including supersoft X-ray sources. Before doing so it is important, however, to compare results obtained with our model grids with results that employed state-of-the-art models for some well-studied objects and to verify that our simplifying assumptions (i.e., LTE and hydrostatic approximation) are justified and that the deduced WD parameters are reasonable. To do that, we converted our model grids to XSPEC table model format and applied them to fit the spectra of two classical SSSs, namely, CAL~$83$ and RX~J0513.9$-$6951. The spectra were obtained by the \textit{Chandra} X-ray observatory using the High Resolution Camera~(HRC-S) and the Low Energy Transmission Grating~(LETG). They are publicly available in the Chandra Grating-Data Archive and Catalog\footnote{\url{http://tgcat.mit.edu}}~\citep[TGCat, ][]{Huenemoerder_etal2011}; see also Table~\ref{tab:log} for the observation log. To increase the signal-to-noise ratio (S/N) the positive and negative first-order spectra were co-added using the \texttt{combine\_grating\_spectra} task of the Chandra Interactive Analysis of Observations~\citep[CIAO,][]{Fruscione_etal2006, CIAO_ascl} package. We also rebinned the spectra to at least $30$~counts per bin in the considered~$19-60\,\AA$~wavelength range. 

For the source CAL~$83,$ we also used the spectrum obtained by the {\it XMM-Newton} Reflection Grating Spectrometer~\citep[RGS,][]{denHerder_etal2001} instrument. It is available in the {\it XMM-Newton} Science Archive\footnote{\url{http://nxsa.esac.esa.int/nxsa-web}} (only the first-order spectrum was used). Table~\ref{tab:log} shows the observation log.

For both sources, due to degeneracy in the WD mass,~$M$, the standard gradient descent method of searching the statistics-minimum and the confidence interval does not allow us to reliably estimate parameter errors. Therefore, the Bayesian Markov chain Monte Carlo (MCMC) approach was used. In particular, we relied on the Bayesian X-ray Analysis~\citep[BXA,][]{Buchner_etal2014, BXA_ascl} package, which allows for the use of the the nested sampling package UltraNest\footnote{\url{https://johannesbuchner.github.io/UltraNest/}}~\citep[][]{Buchner_etal2014, Buchner2019,Buchner2021, UltraNest_ascl} within~XSPEC and allows us to derive the Bayesian evidence and posterior probability distributions. For visualisation of the posterior distributions the \texttt{corner}\footnote{\url{https://corner.readthedocs.io/en/latest}} package~\citep{Foreman-Mackey2016, corner_ascl} was used. The results of the modeling for both objects are discussed in the following.

It should be noted that for low-count spectra the usage of C-statistics~\citep{Cash1979} instead of~$\chi^2$ is recommended; therefore, we used its Xspec implementation (\texttt{cstat}) as likelihood function to determine the best-fit parameters. The parameter uncertainties are derived from the~$0.16$th~and~$0.84$th~quantiles of the posterior distribution. As there is no convenient criterion of estimating the goodness-of-fit (like $\chi^2_{\nu}$) for~\texttt{cstat}, we will give below the statistics value and the number of degrees of freedom.

\begin{table*}
    \caption{{\it Chandra} and {\it XMM-Newton} observations of selected sources.}
\begin{center}
    \begin{tabular}{rrrrrrr}\hline\hline \noalign{\smallskip}
        Name & Instrument & ObsID & Exp. time$^a$ & Start Time (UT) & End Time (UT) & MJD (d)$^b$ \\[0.3em] \hline\noalign{\smallskip}
        CAL~83 & RGS & $0123510101$ & 45.1 & 2000 Apr 23 07:34 & 2000 Apr 23 20:09 & 51657.32 \\
        CAL~83 & HRC-S / LETG & $1900$ & $34.2$ & 2001 Aug 15 16:03 & 2001 Aug 16 02:10 & $52136.67$ \\
        RX~J0513.9$-$6951 & HRC-S / LETG & $5442$ & $25.5$ & 2005 Mar 05 05:40 & 2005 Mar 05 13:18 & $53493.24$ \\
        \hline 
    \end{tabular}
\end{center}
Notes: (a)~-- Exposure time in ks; (b)~-- MJD of the start time.
\label{tab:log}
\end{table*}

\subsubsection{CAL~83}

The prototypical SSS CAL~$83$ was discovered by the \textit{Einstein} observatory in the LMC \citep{1981ApJ...248..925L} and has since been observed by many X-ray observatories, including 
ROSAT \citep{1991A&A...246L..17G}, 
\textit{Beppo}SAX \citep{1998A&A...332..199P}, 
\textit{XMM-Newton} \citep{2001A&A...365L.308P}, 
\textit{Chandra} \citep{2005ApJ...619..517L}, 
and \textrm{NICER} \citep{2022ApJ...932...45O}. 
The X-ray flux of the source is not stable and it sometimes switches to the off-state \citep{1997ASPC..121..730K}; see also \citet{2002A&A...387..944G} and the references therein.
The X-ray pulsations with a period close to $67\,\rm s$ were also discovered by \citet{2014MNRAS.437.2948O} and this period is probably connected with the spin period of the WD. Recently, the available {\it XMM-Newton} X-ray spectra of the source were analysed by \citet{2023MNRAS.522.3472S}.

The optical counterpart of the source is also known as a blue variable star with $V{\approx}17\,\rm mag$~\citep{1984ApJ...286..196C} and its optical spectrum contains prominent Balmer and~\ion{He}{II} emission lines~\citep{1987ApJ...321..745C}. A short review of the optical and ultraviolet (UV) observations of CAL~$83$ can be found in~\citet{2022AJ....164..145S}. In particular, \citet{1998A&A...333..163G} determined the interstellar neutral hydrogen column density $N_{\rm H} = 6.5\times 10^{20}\,{\rm cm}^{-2}$.

Model atmospheres of hot~WDs were used to fit the soft X-ray spectra of CAL~$83$ observed by various X-ray observatories. Spectra of LTE~model atmospheres computed by~\citet{1994A&A...288L..45H} at a fixed $\log g=9$ gave the effective temperature of the hot~WD as $510.5^{+73}_{-7}\,{\rm kK}$ using \textit{Beppo}SAX observations~\citep{1998A&A...332..199P}. 
To reproduce the ROSAT observations, the LTE models were computed by~\citet{2003ARep...47..186I} for several $\log g$~values. However, the poor energy resolution of ROSAT observations did not provide the possibility to limit the surface gravity. The obtained $T_{\rm eff}$ varies from $504\pm17\,{\rm kK}$ at $\log g =8.0$ to $620\pm25\,{\rm kK}$ at $\log g =9.5$, assuming the interstellar column density is fixed at $N_{\rm H} = 6.33\times 10^{20}$\,cm$^{-2}$.

Non-LTE model atmospheres, computed using the publicly available code \texttt{TLUSTY}~\citep{1988CoPhC..52..103H} were used to fit the grating spectra obtained by \textit{XMM-Newton}~\citep{2001A&A...365L.308P} and \textit{Chandra}~\citep{2005ApJ...619..517L}. In the first case, the obtained parameters were very approximate, $T_{\rm eff} \sim 520\,{\rm kK}$ and $\log g \sim 8.5$. However, in the second paper, the model atmospheres were computed specifically for CAL~$83$ case and more reliable results were obtained (see Table~\ref{tab:par}).

 Using our grid with the chemical composition typical for the LMC~($A=0.5$), we fitted simultaneously~{\it Chandra} and XMM~spectra of CAL~$83$. A uniform prior distribution was set for the hydrogen column density, $N_{\rm H}$ (in the range $(1-10)\times 10^{20} \,{\rm cm^{-2}}$), effective temperature, $T_{\rm eff}$ (in the range $100-1000\,{\rm kK}$), white dwarf mass,~$M$ (in range $0.3-1.4\, {\rm M_{\sun}}$), and radius,~$R$ (in range $(2-20)\times 10^8\, {\rm cm}$) was used. We note that we set the strict upper limit for the WD~mass. Another theoretical limitation is based on the fact that the WD~radius must be greater than the cold WD~radius at such a mass \citep[see, e.g.,][]{1972ApJ...175..417N}. Therefore, we further used the $M-R$ relation for cold~WDs as an additional lower limit for a radius at the given WD~mass.

The obtained posterior distribution of fit parameters is presented in Fig.~\ref{fig:corncal83} and Table~\ref{tab:par}. The determined absorption column density,~$N_{\rm H}\sim5.13\times 10^{20}\,\rm cm^{-2}$, is slightly lower than the fiducial value~$(6.5\pm 1)\times 10^{20}\,\rm cm^{-2}$ found previously by \citet{1998A&A...333..163G}, who analyzed the UV spectra of CAL~$83$ and RX~J0513.9$-$6951. Therefore, we performed the additional modelling assuming fixed $N_{\rm H}=6.5\times 10^{20} \,{\rm cm^{-2}}$ and $5.5\times 10^{20} \,{\rm cm^{-2}}$ (average column density in the direction of the LMC). These results are also presented in Table~\ref{tab:par}. It is clearly seen from the statistics value that the quality of these fits are worse, with the WD~mass approaching the hard upper limit.

The comparison of the observed and model {\it Chandra} spectrum, corresponding to a model with $N_{\rm H}$ as a free parameter, is shown in Fig.~\ref{fig:fit}, left panel. Despite using the LTE~model atmospheres, the parameters found are very close to those found by~\citet{2005ApJ...619..517L} (see Table~\ref{tab:par}). Our LTE~atmosphere models thus yield the same SSS~parameters with a similar accuracy as the non-LTE model spectra. However, our grid covers a much larger parameter space, which also means it may be applied to other sources.

\begin{figure}
\centering
\includegraphics[width=1.0\columnwidth]{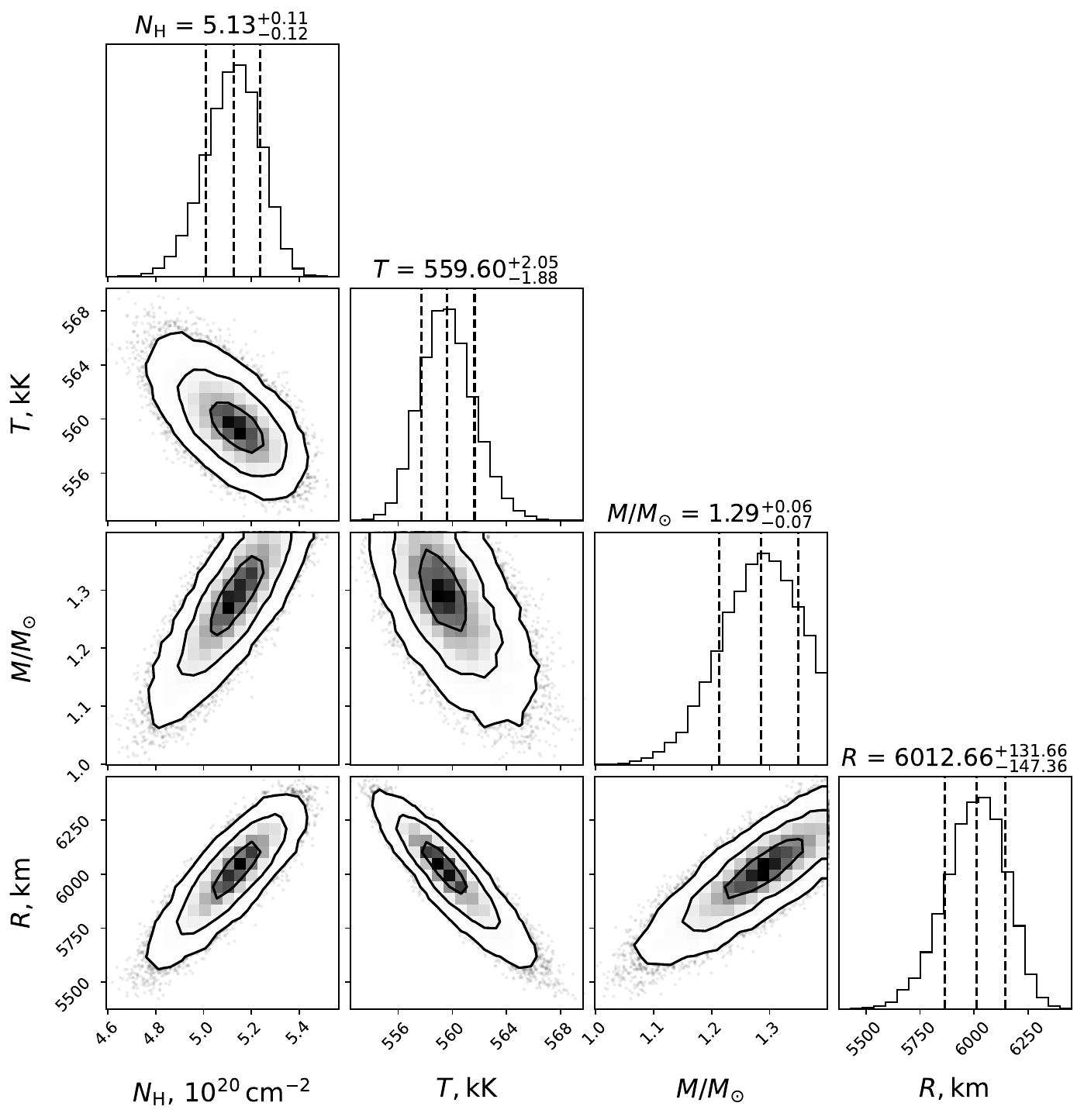}
\caption{Corner plot of the posterior distribution for parameters of CAL~$83$. The 2D contours correspond to~$39.3\%$, $86.5\%$, and~$98.9\%$ confidence levels. The histograms show the normalized 1D distributions for a given parameter derived from the posterior samples. The best-fit parameter values are presented in Table~\ref{tab:par}. A Gaussian prior for the hydrogen column density $N_H$ was used, with $\mu = 6.5\times 10^{20}\,\rm cm^{-2}$ and $\sigma = 1.0\times 10^{20}\,{\rm cm}^{-2}$. 
}
\label{fig:corncal83}
\end{figure}

\begin{table*}
    \caption{Parameters of the investigated SSSs.} 
\begin{center}
    \begin{tabular}{l||lllll|lll}
    Parameter  &  CAL~$83^a$ & CAL~$83^a$ & CAL~$83^a$ &  CAL~$83^b$ & CAL~$83^a$(bb) & RX~J$0513^a$  & RX~J$0513^c$ & RX~J$0513^a$(bb) \\
    \hline \\[-0.9em] 
        $N_{\rm H}, 10^{20}\,\rm cm^{-2}$  & $6.5^d$   & $5.5^d$   & $5.13 \pm_{0.12}^{0.11}$   & $6.5 \pm 1.0$ & $5.13^d$  & $5.40 \pm_{0.10}^{0.09}$  & $5.94 \pm_{0.4}^{0.47}$ & $5.4^d$  \\[0.3em]
        $T_{\rm eff},\,\rm kK$  & $533 \pm 2$  & $556 \pm 1$  & $560 \pm 2$ & $550 \pm 25$ & $540 \pm_{2}^{3}$ & $629 \pm_{4}^{6}$   & $594 \pm_{7}^{10}$  & $684 \pm 4$  \\[0.3em]
        $M/M_{\sun}$ & $1.39 \pm 0.01$  & $1.39 \pm_{0.02}^{0.01}$  & $1.29 \pm_{0.07}^{0.06}$ & $1.3 \pm 0.3$ & & $1.33 \pm_{0.07}^{0.05}$ & $1.0 \pm 0.2 $ & \\[0.3em]
        $R\,^e,\,10^8$\,\rm cm  & $7.8 \pm 0.1$ & $6.3 \pm 0.1$ & $6.0 \pm 0.1$ & $7.0 \pm 0.7$ & $8.6 \pm 0.2$ & $7.0 \pm 0.2$ & $10 \pm 2$     & $6.5 \pm 0.1$\\[0.3em]
        $L^e, 10^{37}\,\rm erg\,s^{-1}$  & $3.5 \pm 0.1 $ 
        & $2.7 \pm 0.1 $ & $2.5 \pm 0.1 $ & $3.5 \pm 1.2$ 
        & $4.4 \pm 0.2$ & $5.4 \pm 0.4$    & $7.5  \pm 2$ 
        & $6.5 \pm 0.3$ \\[0.3em]
        $\log g$  & $8.48 \pm 0.02$ & $8.66 \pm 0.01$ & $8.67 \pm 0.03$ & $8.5 \pm 0.1$ &  & $8.56 \pm_{0.04}^{0.03}$ & $8.4 \pm_{0.15}^{0.04}$ & \\[0.3em]
        $\Delta \log g$ & $0.75 \pm 0.02$ & $0.86 \pm 0.01$ & $0.86 \pm 0.03$ &$\approx 0.7$ & & $0.55 \pm_{0.04}^{0.03}$ & $0.49 \pm_{0.15}^{0.04}$ & \\[0.5em]
        \multirow{2}{*}{$\texttt{cstat}\,\,{\rm (dof)}^f$} & $7005.47$ & $6823.29$ & $6811.52$ & & $6438.81$ & $2114.14$ & & $2175.42$\\[0.3em]
        & $(3807)$ & $(3807)$ & $(3806)$ & & $(3808)$ & $(575)$ & & $(577)$ \\[0.3em]
        \hline 
    \end{tabular}
\end{center}
Notes: (a)~-- parameters obtained in this work; (b)~-- parameters obtained by~\citet{2005ApJ...619..517L}, (c)~-- parameters obtained by~\citet{2003ARep...47..186I, 2003ARep...47..197S}; (d)~--- hydrogen column density $N_H$  is fixed; (e)~-- the distance to the~LMC is assumed to be~$50\,\rm kpc$ \citep{2019Natur.567..200P}; (f)~--  C-statistics and the number of degrees of freedom for the best fit found using Bayesian analysis.
\label{tab:par}
\end{table*}

\begin{figure*}
\centering
        \begin{minipage}{0.49\linewidth}
        \center{\includegraphics[width=1.0\linewidth]{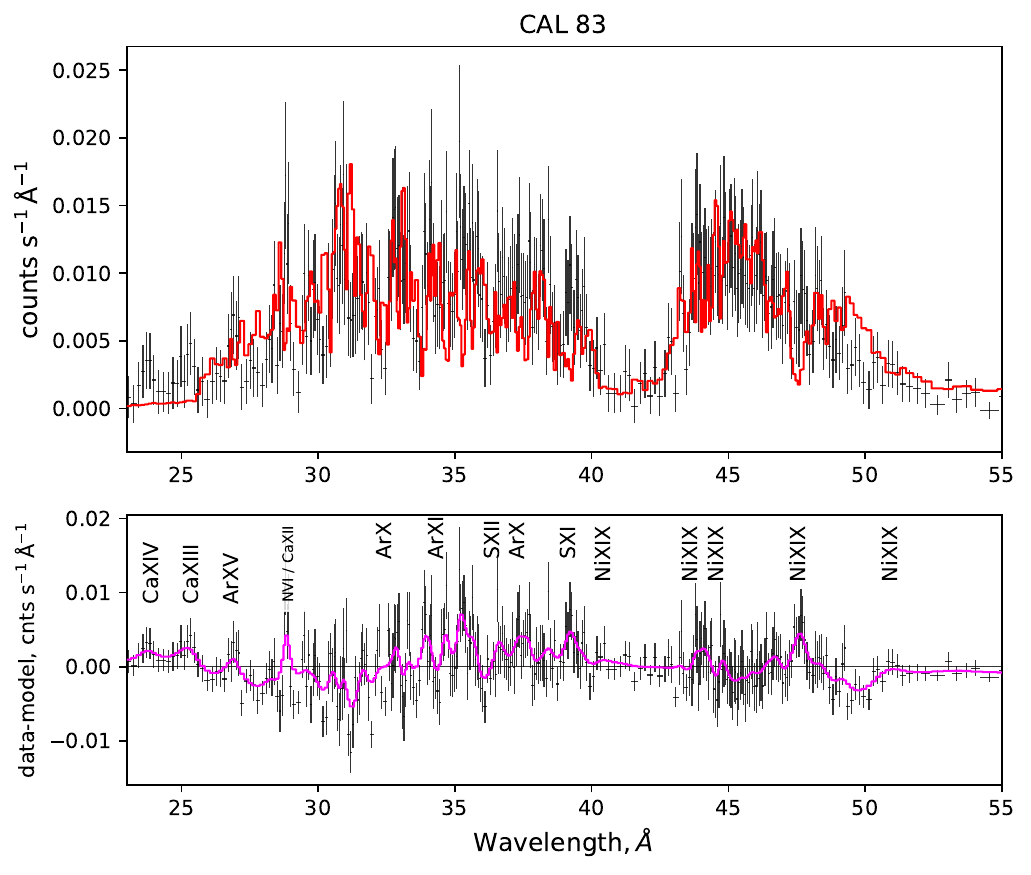}}
        \end{minipage}
        \begin{minipage}{0.49\linewidth}
        \center{\includegraphics[width=1.0\linewidth]{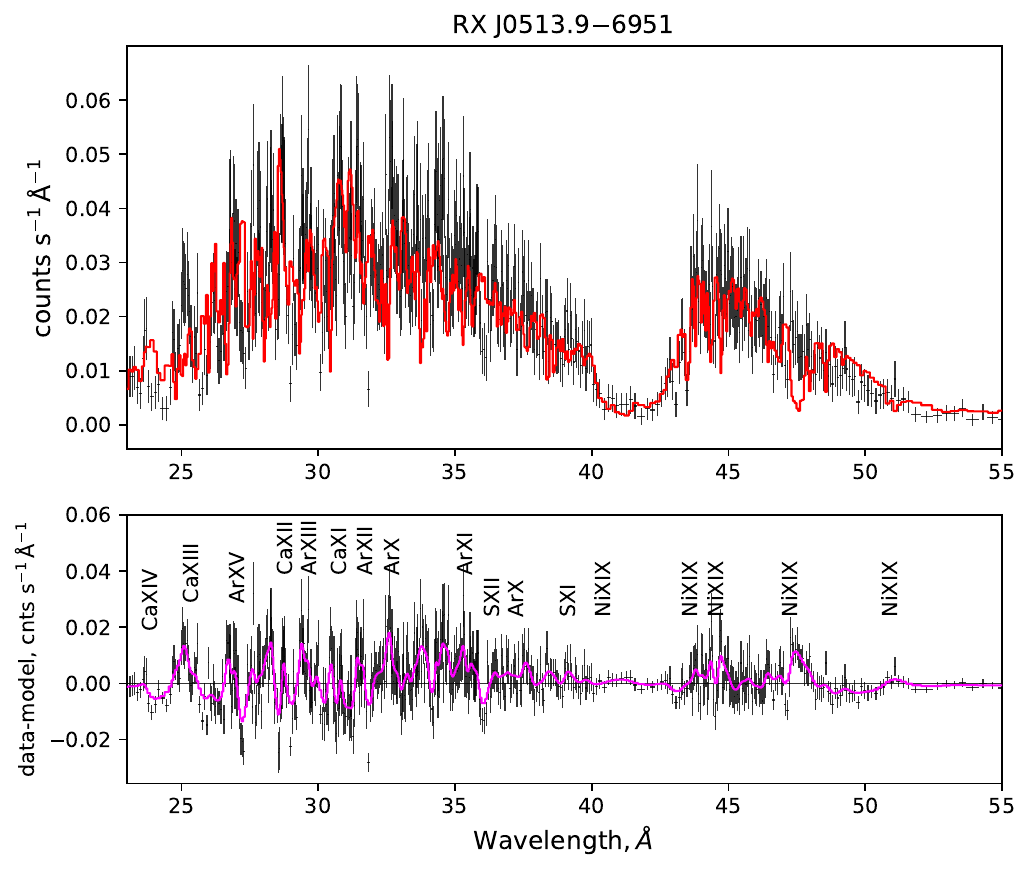}}
        \end{minipage}
\caption{Best fits of CAL~$83$ (left) and RX~J0513.9$-$6951 (right) spectra (top). The observed spectra (\textit{Chandra}, black curves) and the model fits (red curves) are shown. Residuals and the smoothed residuals (magenta curves) of the fits (bottom). Any possible identifications of the most prominent residual features as resonance lines of highly charged ions of some heavy elements are indicated. The best-fit parameters are presented in Table~\ref{tab:par}.}
\label{fig:fit}
\end{figure*}

The spectrum of CAL~83 observed by~\textit{Beppo}SAX was fitted with a blackbody model by \citet{1998A&A...332..199P} and the quality of the fit was even better compared to the non-LTE model atmospheres fits. Here, we also fitted the~XMM~and~\textit{Chandra} spectra with the blackbody model. Fitting with the free~$N_{\rm H}$ yields larger values of~$N_{\rm H}$ ($\approx 13.3\times 10^{20}$\,cm$^{-2}$) and a highly non-physically large WD~radius ($\approx 13.4\times 10^{9}$\,cm), but a lower $T_{\rm eff}$ ($\approx 358$\,kK). In this case the derived bolometric luminosity exceeds the Eddington luminosity for a solar mass object, $\approx 2 \times 10^{39}$\,erg\,s$^{-1}$. 
However, if we fix the interstellar absorption at a value found by the model atmosphere fitting, the best blackbody fit yields results that do not differ significantly from those of the best-fit model atmosphere (see Table~\ref{tab:par}). We note that formal fit quality is actually slightly better for the blackbody model, however, we emphasize that both models are statistically acceptable; moreover, the model atmosphere fits yield satisfactory results 
without the need to fix the hydrogen column density, unlike the blackbody fitting.

\begin{table}
    \caption{Parameters of CAL\,83 after fitting by the models with various chemical compositions.} 
\begin{center}
    \begin{tabular}{llll}
    Parameter  &  $A=0.1$ & $A=1$ &  $A=1$ \\
    \hline \\[-0.9em] 
        $N_{\rm H}, 10^{20}\,\rm cm^{-2}$  & $5.46 \pm 0.08$   & $3.0\pm 0.1$ &  $5.1^a$ \\[0.3em]
        $T_{\rm eff},\,\rm kK$  & $508 \pm 2$  & $716 \pm 6$  &  $586 \pm 2$ \\[0.3em]
        $M/M_{\sun}$ & $1.39 \pm 0.01$  & $1.35 \pm 0.01$  & $1.04 \pm 0.01$\\[0.3em]
        $R\,^b,\,10^8$\,\rm cm  & $8.01 \pm 0.14$ & $2.3 \pm 0.07$ & $5.17 \pm 0.05$\\[0.3em]
        $L^b, 10^{37}\,\rm erg\,s^{-1}$  & $3.04 \pm 0.1 $ & $1.0 \pm 0.1 $ & $2.25 \pm 0.1 $\\[0.3em]
        $\log g$  & $8.46 \pm 0.02$ & $9.53 \pm 0.03$ & $8.72 \pm 0.02$\\[0.3em]
        $\Delta \log g$ & $0.82 \pm 0.02$ & $1.29 \pm 0.03$ & $0.83 \pm 0.02$\\[0.5em]
        \multirow{2}{*}{$\texttt{cstat}\,\,{\rm (dof)}^c$} & $6820.23$ & $6172.02$ & 6995.9 \\[0.3em]
        & $(3806)$ & $(3806)$  & $(3807)$\\[0.3em]
        \hline 
    \end{tabular}
\end{center}
 Notes: (a)~hydrogen column density $N_H$  is fixed; (b)~distance to the~LMC is assumed to be~$50\,\rm kpc$ \citep{2019Natur.567..200P}; (c)~C-statistics and the degrees of freedom for the best fit found with Bayesian analysis.
\label{tab:sol}
\end{table}

We also performed a fitting of the joint {\it Chandra} and {\it XMM-Newton} grating spectra using models with the solar ($A=1$) chemical composition and models with heavy elements reduced by a factor of ten ($A=0.1$). The results are presented in Table\,\ref{tab:sol}.
The main conclusion is that there are remarkable differences between the best fits for the models with different chemical composition. 
The models with ten times reduced heavy element abundances ($A=0.1$) produce the fit with a slightly decreased effective temperature, slightly increased photospheric radius and increased interstellar absorption. It was expected because the model spectra
with $A=0.1$ are slightly harder than those with $A=0.5$
(see Figure\,\ref{fig:500gA}, bottom panel).
Therefore, a model with a reduced temperature is needed to reproduce the observed spectrum. On the other hand, fitting using the models with solar heavy element abundances~($A=1$) demonstrates more significant differences compared to the model spectra with~$A=0.5$ (see Table\,\ref{tab:sol}). Formally, the fit quality is better with $N_{\rm H}$ as a free parameter. However, the obtained~$N_{\rm H}$ is too small, the temperature is too high and the radius is too small as well. In fact, the radius approached the low limit boundary for cold WD~radius. 
Therefore, we fixed the interstellar absorption at the value obtained for $A=0.5$. The resulting fit is much closer to the fit obtained for~$A=0.5$ (see Table\,\ref{tab:sol}) and it demonstrates a slightly increased temperature and a slightly decreased photospheric radius. All the fits 
($A=0.5$, free $N_{\rm H}$; $A=0.1$, free $N_{\rm H}$; and $A=1$, fixed $N_{\rm H}$) are not distinguishable from the statistical point of view.

\subsubsection{RX~J0513.9--6951}

This recurrent SSS was discovered in ROSAT data by \cite{1993A&A...270L...9S}, and the corresponding optical counterpart was almost immediately identified by \cite{1993A&A...278L..39P}. The most interesting property of the source is its X-ray off-states, accompanied by optical brightening during these intervals~\citep{1996ApJ...470.1065S}. It is most likely (see the cited work) that the source's luminosity during X-ray off-states is so high that it leads to a significant expansion of the envelope radius and the effective temperature drops, making the emergent radiation too soft to be observed in X-rays.

The X-ray spectra of RX~J0513.9$-$6951 have been described by model atmosphere spectra only a couple of times. A low-resolution ROSAT spectrum was fitted with simple LTE~model atmospheres~\citep{2003ARep...47..186I}.
Fitting a grating \textit{Chandra} spectrum with the same models was less successful, and a two-component model was used to enhance the fit quality~\citep{2007AdSpR..40.1294B}. Numerous \textit{XMM-Newton} spectra obtained by the Resolution Grating Spectrometer~(RGS)~\citep{2005MNRAS.364..462M} were not fitted with the model atmosphere spectra.

\begin{figure}
\centering
\includegraphics[width=1.0\columnwidth]{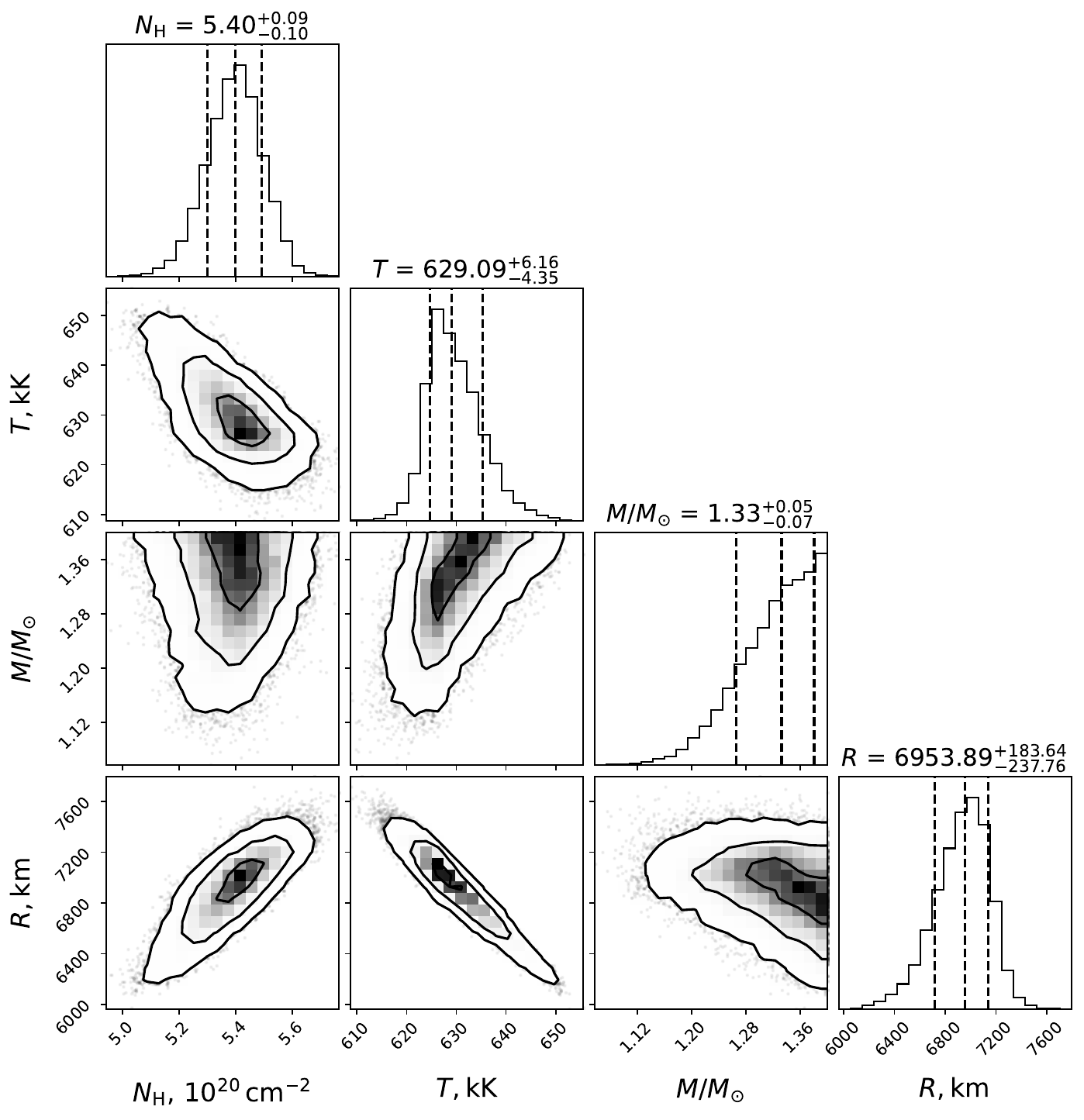}
\caption{Corner plot of the posterior distribution for parameters of RX~J0513.9$-$6951. The 2D contours correspond to~$39.3\%$, $86.5\%$, and~$98.9\%$ confidence levels. The histograms show the normalized 1D distributions for a given parameter derived from the posterior samples. The best-fit parameter values are presented in Table~\ref{tab:par}.
}
\label{fig:cornrx513}
\end{figure}

The results of fitting the~\textit{Chandra} spectrum with the new models described above are presented in Table~\ref{tab:par} and the spectrum is shown in Fig.~\ref{fig:fit} (right panels). The corresponding posterior distribution of fit parameters is shown in Fig.~\ref{fig:cornrx513}.
Similarly to CAL~83, we assumed uniform prior distributions for the hydrogen column density, $N_{\rm H}$ (range $(1-10)\times 10^{20} \,{\rm cm^{-2}}$); effective temperature, $T_{\rm eff}$ (range $100-1000\,{\rm kK}$); the white dwarf mass,~$M$ (range $0.3-1.4\, {\rm M_{\sun}}$); and radius~$R$ (range $(2-20)\times 10^8\, {\rm cm}$).
It should be noted that the obtained hydrogen column density,~$N_{\rm H}$, posterior value is consistent with the result from~\cite{1998A&A...333..163G}; namely, $(5.5\pm1.0)\times 10^{20}\,{\rm cm}^{-2}$. Also, the found values of $T_{\rm eff}$ and $\log g$ are close to those obtained from the ROSAT spectrum~\citep{2003ARep...47..186I}, shown in Table~\ref{tab:par}. 

We also fitted the spectrum of RX\,J0513.9$-$6951 with a blackbody model. The fit with a free $N_{\rm H}$ has the same features like the CAL\,83 fit: $N_{\rm H}$ value and WD~radius are significantly larger than in the model atmosphere fit, and the blackbody temperature is much lower. The fit with the fixed~$N_{\rm H}$ also yields results that are relatively close to those of the model atmosphere fit (see Table~\ref{tab:par}). However, comparing the blackbody fit with the fixed~$N_{\rm H}$ and the model atmosphere fit with the free~$N_{\rm H}$ using the Bayes factor showed that the spectrum is slightly better described by the model atmosphere spectra than by the blackbody.

\subsubsection{Emission component importance}

It is widely known that the spectra of highly inclined~SSSs are dominated by emission lines~\citep[see, e.g.,][]{2013A&A...559A..50N}. Possibly, the same or similar emission lines have to exist in the spectra of CAL~$83$ and RX~J0513.9$-$6951. 
For example, three emission features between~$24$~and~$28$~\AA\ are seen in the observed spectra of both sources.
Most probably they are resonance lines of \ion{Ca}{XIV} ($\sim 24$~\AA), \ion{Ca}{XIII} ($\sim 25.2$~\AA), and \ion{Ar}{XV} ($\sim 26.8$~\AA). Observed flux excesses, visible at longer wavelengths, may also represent emission lines of highly charged ions of the same or similar elements. We assume that these emission line components are thermal emission of the line-driven winds from the hot~WD surface. The presence of the emission component affects the accuracy of spectrum fitting and it has to be included in further modeling.

The lower panels in Fig.~\ref{fig:fit} show
the residuals between the observed spectra and the hot~WD model atmosphere spectra, with the parameters listed in Table~\ref{tab:par}. The most prominent emission features can be identified with the resonance lines of highly charged ions of heavy elements. The list of identified lines is presented in Table~\ref{tab:lines}. The line identification was made using CHIANTI~line database~\citep{1997A&AS..125..149D}.

We note, however, that the significance of the interpreted emission lines is not high enough. Formally, we can improve the quality of the fits by including a few~$(5-6)$ emission lines. This procedure formally reduces the reduced $\chi^2$ (by ${\sim}\,0.2$), but does not significantly change the WD~atmosphere parameters.

\begin{table}
    \caption{Lines likely corresponding to observed emission features.} 
\begin{center}
    \begin{tabular}{l|ccc}
 Ion  & $\lambda$ (\AA)  & $gf$ & $E_{\rm low}$\,(eV)  \\
    \hline
    \hline \\[-0.9em] 
Ca XIV  &   24.09 & 2.39  & 0.00 \\
        &   24.13 & 3.58  & 0.00 \\ \hline
Ca XIII &   25.53 & 2.35  & 0.00 \\ \hline
Ar XV   &   26.62 & 2.91  & 0.00 \\
        &   26.66 & 2.24  & 0.00 \\
        &   26.71 & 3.93  & 0.00 \\ \hline
Ca XII  &   28.48 & 1.39  & 0.00 \\ \hline
N VI    &   28.79 & 0.66  & 0.00 \\
Ca XII  &   28.86 & 0.76  & 3.73 \\ \hline
Ar XIII &   29.32 & 1.12  & 2.72 \\
        &   29.32 & 0.91  & 0.00 \\
        &   29.35 & 2.04  & 1.22 \\
        &   29.37 & 3.81  & 2.72 \\ \hline
Ca XI   &   30.45 & 2.34  & 0.00 \\
        &   30.45 & 2.68  & 0.00 \\ \hline
Ar XII  &   31.35 & 2.32  & 0.00 \\
        &   31.39 & 3.47  & 0.00 \\ \hline
Ar X    &   32.45 & 0.14  & 0.00 \\
        &   32.61 & 0.44  & 0.00 \\ \hline
Ar XI   &   34.10 & 0.81  & 0.00 \\
        &   34.24 & 1.80  & 0.00 \\
        &   34.52 & 1.12  & 1.79 \\ \hline
Ar XI   &   35.37 & 1.08  & 0.00 \\ \hline
S XII   &   36.56 & 4.10  & 1.63 \\ \hline
Ar X    &   37.43 & 2.61  & 0.00 \\
        &   37.48 & 1.89  & 0.00 \\ 
        &   38.23 & 1.46  & 0.00 \\ \hline
S XI    &   39.24 & 2.73  & 1.54 \\
        &   39.24 & 2.00  & 0.00 \\
        &   39.30 & 4.10  & 0.65 \\
        &   39.32 & 3.59  & 1.54 \\ \hline
C V     &   40.27 & 0.65  & 0.00 \\
Ni XIX  &   40.60 & 3.08  & 0.00 \\ \hline
Ni XIX  &   43.79 & 1.75  & 0.00 \\ 
        &   44.73 & 3.83  & 0.00 \\ \hline
S IX    &   47.43 & 1.89  & 0.00 \\
Ni XIX  &   47.43 & 8.38  & 0.00 \\
        &   47.52 & 8.99  & 0.00 \\
        &   47.55 & 3.83  & 0.00 \\
        &   47.65 & 5.91  & 0.00 \\
        &   47.73 & 5.97  & 0.00 \\ \hline
Ni XIX  &   51.09 & 5.01  & 0.00 \\ \hline
    \end{tabular}
\end{center}    
Notes: Lines probably corresponding to one feature are separated by horizontal lines. The wavelengths, $gf$-factors, and the excitation energy of the lower levels $E_{\rm low}$ are also presented. Some close lines of the same ion and the low energy level have been merged.
\label{tab:lines}
\end{table}

\section{Discussion}
\label{sec:discussion}

Our spectral fitting showed that the WD masses in the investigated sources are high ($> 1.2\,M_\odot$) and tend to the maximum possible values for WDs. However, the derived radii are significantly larger than radii of cold WDs with these masses. It is, in fact, expected because radii of single~WDs with hydrogen or helium envelopes and finite temperatures are larger than cold WD radii~\citep[see, e.g.,][]{2001PASP..113..409F}. On the other hand, the difference in the radius between the cold WDs and the WDs with the hot envelopes is minimal for high-mass WDs. 
\begin{figure}
\centering
\includegraphics[width=0.95\columnwidth]{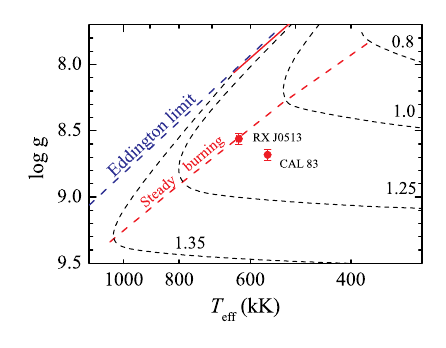}
\caption{Positions of the investigated~SSSs in the $T_{\rm eff} - \log g$ plane. Model dependencies for various WD~masses~\citep{2007ApJ...663.1269N} are shown by black dashed curves with indicated WD~masses (in solar masses). The boundaries of the stable thermonuclear burning band are shown by the solid (upper boundary) and the dashed (lower boundary) red lines. The Eddington limit for solar H/He abundances is shown by the blue dashed line. 
}
\label{fig:glogT}
\end{figure}

The envelope temperatures with thermonuclear burning on the WD~surface in SSSs are, however, significantly higher than those of thick envelopes of isolated WDs. Therefore, we expect that the photospheric radius of a hot~WD in SSSs could be significantly larger than the radius of a cold~WD with the same mass, and comparison with the computed models is potentially important. 
Models of WD~envelopes with hydrogen thermonuclear burning have been computed by several groups~\citep[see, e.g.,][]{2007ApJ...663.1269N,2013ApJ...777..136W}. The numerical dependencies of the WD~radii on the surface effective temperatures for a few WD~masses are presented by~\citet{2007ApJ...663.1269N}. We plotted these dependencies in the $T_{\rm eff} - \log g$ plane (see Fig.~\ref{fig:glogT}) together with the positions of CAL~$83$ and RX~J0513.9$-$6951. We took the fits obtained for free $N_{\rm H}$.
It is clearly seen that the WD masses in the SSSs derived from the model dependences are less than we obtained from the spectral fitting, and correspond to about $1.1 - 1.15\, M_\odot$ for both sources. 

\begin{figure}
\centering
\includegraphics[width=0.99\columnwidth]{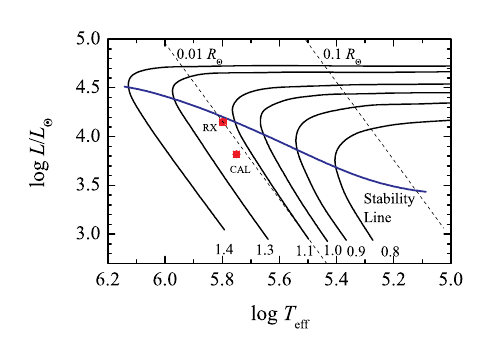}
\includegraphics[width=0.99\columnwidth]{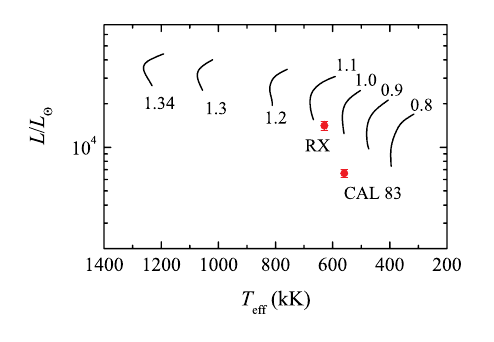}
\caption{Positions of the investigated SSSs in the $T_{\rm eff} - L/L_\odot$ planes. \textit{Top panel:} Model dependencies for various WD~masses were taken from~\citet{1982ApJ...259..244I}. Steady-state 
thermonuclear burning on WD surfaces is possible
above the stability line only. \textit{Bottom panel:} Model dependencies were taken from~\citet{2013ApJ...777..136W}. Only the model curves with steady-state thermonuclear burning are shown.
}
\label{fig:LT} 
\end{figure}

Another theoretical constraint can be obtained from the cooling tracks in the~$\log T_{\rm eff} - \log L/L_\odot$~plane,
where $L$ is a luminosity.
Such tracks computed (among others) by~\citet{1982ApJ...259..244I} are presented in Fig.~\ref{fig:LT}\,(top panel; also see Figure~2 in the cited paper), with the added positions of CAL~$83$ and RX~J0513.9$-$6951. 

The positions of the SSSs correspond to almost the same WD masses, $1.15 -1.2\, M_\odot$. 
We also compared the sources' positions on this plane with 
the model predictions by~\citet{2013ApJ...777..136W} (Fig.~\ref{fig:LT}, bottom panel).
In this case, the positions correspond to the cooling tracks with a slightly lower mass, namely, $1.05-1.1\, M_\odot$.

These results confirm a conclusion made previously \citep{2003ARep...47..197S}, which posits that RX~J0513.9$-$6951 lies in the steady-state thermonuclear burning band, as also seen in Figure~5 in \citet{2003ARep...47..186I}. Therefore, the off-states of this SSS are most probably connected with the photospheric radius expansion during periods of increased mass-accretion rates. At the same time, the luminosity of the hot~WD in CAL~$83$ is significantly lower than the predicted one for the steady state burning band. It means that this source lies below the stable thermonuclear burning band, as found earlier~\citep{2003ARep...47..197S}. Therefore, thermonuclear burning most likely arises episodically and the X-ray off-states are connected with the cessation of thermonuclear burning~\citep{1995A&A...304..227K,1998A&A...331..328K}. In the  cited work, there is a statement that a relatively short off-states duration in CAL~$83$ ($<100 \rm\,days$) can exist if the~WD is quite massive, $M > 1.2\,M_\odot$. Our estimation of the WD~mass in CAL~$83$ derived from the spectral fit is in agreement with this condition.

\section{Summary}
\label{sec:summary}

In this paper, we present a new extended set of hot WD~model spectra. The corresponding model atmospheres were computed under LTE and hydrostatic equilibrium assumptions. The list of spectral lines presented by CHIANTI~collaboration was used. The novel thing in contrast with our previous computations is taking into account the photoionization from the excited atomic levels. 
The set was computed for three chemical compositions. The solar hydrogen-helium mix was used for all models, but the heavy element abundances were taken to be equal to the solar one~$(A=1)$, half of the solar~$(A=0.5)$, and one-tenth of the solar~$(A=0.1)$. These abundances correspond to the Milky Way disc, the LMC, and the SMC. The grid covers $T_{\rm eff} = 100 - 1000\,{\rm kK}$ in steps of $25\,{\rm kK}$, and eight values of surface gravity, measured from the limited possible surface gravity $g_{\rm Edd}$, $\Delta \log g = \log g - \log g_{\rm Edd} = 0.1, 0.2, 0.4, 0.6, 1.0, 1.4, 1.8,$ and $2.2$. 
This model spectra set is designed to fit the observed soft X-ray spectra of SSSs and it was implemented into~XSPEC\footnote{\red{\url{https://github.com/HEASARC/xspec_localmodels/tree/master/sss_atm}}}. 

We used the calculated model grid to interpret the grating spectra of two bright~SSSs in the LMC, CAL~$83$~(\textit{Chandra}/LETGS and \textit{XMM-Newton}/RGS spectra), and RX~J0513.9$-$6951~(one \textit{Chandra}/LETGS spectrum). 
We found that best-fit WD~effective temperature values are in agreement with the values obtained in previous investigations. The values of~$T_{\rm eff}$ and~$\log g$ obtained for CAL~$83$ using detailed non-LTE model atmospheres are also in good agreement with our results. Therefore, we conclude that our simplified LTE~model atmospheres can be used for the analysis of the X-ray spectra of SSSs with good results despite their limitations. In fact, there are no publicly available sets of non-hydrostatic model atmospheres and non-LTE atmosphere models were computed only in a narrow range of input parameters, so the presented model set is actually the first of its kind to be released to general community.

The WD parameters found for the investigated~SSSs confirm that RX~J0513.9$-$6951 lies in the stable nuclear burning strip, while CAL~$83$ lies below this strip, and thermonuclear burning on the WD surface is episodic. Theoretically, a relatively short observed duration between burning episodes is possible if the~WD~is massive enough, $M > 1.2 M_\odot$. Using published model $T_{\rm eff} - \log g$ and $\log T_{\rm eff} - \log L/L_\odot$ dependencies, we estimated the WD mass in CAL~$83$ 
to be confined within $1.05-1.2\,M_\odot$, which is in marginal agreement with this theoretical prediction. The WD~mass estimation for RX~J0513.9$-$6951 gives a more narrow range, 
$1.1-1.2\,M_\odot$. 

A notable contradiction between the WD masses formally obtained from the spectral fits and from the sources' positions on the theoretical $T_{\rm eff} - \log g$ and $T_{\rm eff} - \log L/L_\odot$ dependencies can be pointed out.
The masses from the spectral fits are high ($\approx 1.29\,M_\odot$ for CAL\,83) and poorly constrained, and tend to the maximum possible WD mass 1.4\,$M_\odot$ for RX\,J0513.9$-$6951. On the contrary, the masses from the model tracks lie within $1.05-1.2\,M_\odot$. 

Although we could also suggest that the model tracks need to be revised, the simplest solution is that the spectral fits provide poor accuracy in determining WD~masses.
The main reason is that we can observe only a relatively narrow soft X-ray part of the general spectral energy distribution of SSSs. Moreover, the soft X-ray part of the spectra is heavily influenced by interstellar absorption. It means that we had to introduce limitation on the $N_{\rm H}$ value from optical and UV observations, and set the limit of the WD~radius, which cannot be smaller than the radius of a cold WD with a given mass.
Perhaps, the development of the observations and the models will allow us to obtain basic SSS parameters, including the chemical composition, only from soft X-ray observations without any additional conditions. However, this is a matter that will be tackled in the distant future.
Finally, we conclude that the presented model spectra set could be useful for interpreting both the existing 
grating spectra of~SSS, obtained by \textit{Chandra}
and \textit{XMM-Newton},
and new observations of SSSs by eROSITA. 

\begin{acknowledgements}  
We express our deep gratitude to the referee for their comments, which significantly improved this article.
VFS and AST thank the Deutsche Forschungsgemeinschaft (DFG) for financial support (grants WE 1312/59-1 and WE 1312/56-1, respectively). 
\end{acknowledgements}

\bibliographystyle{aa}
\bibliography{sss}

\begin{thebibliography}{86}
\expandafter\ifx\csname natexlab\endcsname\relax\def\natexlab#1{#1}\fi

\bibitem[{{Arnaud} {et~al.}(1999){Arnaud}, {Dorman}, \& {Gordon}}]{XSPEC_ascl}
{Arnaud}, K., {Dorman}, B., \& {Gordon}, C. 1999, {XSPEC: An X-ray spectral
  fitting package}, Astrophysics Source Code Library, record ascl:9910.005

\bibitem[{{Arnaud}(1996)}]{Arnaud1996}
{Arnaud}, K.~A. 1996, in Astronomical Society of the Pacific Conference Series,
  Vol. 101, Astronomical Data Analysis Software and Systems V, ed. G.~H.
  {Jacoby} \& J.~{Barnes}, 17

\bibitem[{{Buchner}(2016{\natexlab{a}})}]{BXA_ascl}
{Buchner}, J. 2016{\natexlab{a}}, {BXA: Bayesian X-ray Analysis}, Astrophysics
  Source Code Library, record ascl:1610.011

\bibitem[{{Buchner}(2016{\natexlab{b}})}]{UltraNest_ascl}
{Buchner}, J. 2016{\natexlab{b}}, {UltraNest: Pythonic Nested Sampling
  Development Framework and UltraNest}, Astrophysics Source Code Library,
  record ascl:1611.001

\bibitem[{{Buchner}(2019)}]{Buchner2019}
{Buchner}, J. 2019, \pasp, 131, 108005

\bibitem[{{Buchner}(2021)}]{Buchner2021}
{Buchner}, J. 2021, The Journal of Open Source Software, 6, 3001

\bibitem[{{Buchner} {et~al.}(2014){Buchner}, {Georgakakis}, {Nandra}, {Hsu},
  {Rangel}, {Brightman}, {Merloni}, {Salvato}, {Donley}, \&
  {Kocevski}}]{Buchner_etal2014}
{Buchner}, J., {Georgakakis}, A., {Nandra}, K., {et~al.} 2014, \aap, 564, A125

\bibitem[{{Burwitz} {et~al.}(2007){Burwitz}, {Reinsch}, {Greiner}, {Rauch},
  {Suleimanov}, {Walter}, {Mennickent}, \& {Predehl}}]{2007AdSpR..40.1294B}
{Burwitz}, V., {Reinsch}, K., {Greiner}, J., {et~al.} 2007, Advances in Space
  Research, 40, 1294

\bibitem[{{Carrera} {et~al.}(2008){Carrera}, {Gallart}, {Aparicio}, {Costa},
  {M{\'e}ndez}, \& {No{\"e}l}}]{2008AJ....136.1039C}
{Carrera}, R., {Gallart}, C., {Aparicio}, A., {et~al.} 2008, \aj, 136, 1039

\bibitem[{{Cash}(1979)}]{Cash1979}
{Cash}, W. 1979, \apj, 228, 939

\bibitem[{{CIAO Development Team}(2013)}]{CIAO_ascl}
{CIAO Development Team}. 2013, {CIAO: Chandra Interactive Analysis of
  Observations}, Astrophysics Source Code Library, record ascl:1311.006

\bibitem[{{Cowley} {et~al.}(1984){Cowley}, {Crampton}, {Hutchings}, {Helfand},
  {Hamilton}, {Thorstensen}, \& {Charles}}]{1984ApJ...286..196C}
{Cowley}, A.~P., {Crampton}, D., {Hutchings}, J.~B., {et~al.} 1984, \apj, 286,
  196

\bibitem[{{Cowley}(1971)}]{1971Obs....91..139C}
{Cowley}, C.~R. 1971, The Observatory, 91, 139

\bibitem[{{Crampton} {et~al.}(1987){Crampton}, {Cowley}, {Hutchings},
  {Schmidtke}, {Thompson}, \& {Liebert}}]{1987ApJ...321..745C}
{Crampton}, D., {Cowley}, A.~P., {Hutchings}, J.~B., {et~al.} 1987, \apj, 321,
  745

\bibitem[{{Del Zanna} {et~al.}(2021){Del Zanna}, {Dere}, {Young}, \&
  {Landi}}]{2021ApJ...909...38D}
{Del Zanna}, G., {Dere}, K.~P., {Young}, P.~R., \& {Landi}, E. 2021, \apj, 909,
  38

\bibitem[{{den Herder} {et~al.}(2001){den Herder}, {Brinkman}, {Kahn},
  {Branduardi-Raymont}, {Thomsen}, {Aarts}, {Audard}, {Bixler}, {den Boggende},
  {Cottam}, {Decker}, {Dubbeldam}, {Erd}, {Goulooze}, {G{\"u}del}, {Guttridge},
  {Hailey}, {Janabi}, {Kaastra}, {de Korte}, {van Leeuwen}, {Mauche},
  {McCalden}, {Mewe}, {Naber}, {Paerels}, {Peterson}, {Rasmussen}, {Rees},
  {Sakelliou}, {Sako}, {Spodek}, {Stern}, {Tamura}, {Tandy}, {de Vries},
  {Welch}, \& {Zehnder}}]{denHerder_etal2001}
{den Herder}, J.~W., {Brinkman}, A.~C., {Kahn}, S.~M., {et~al.} 2001, \aap,
  365, L7

\bibitem[{{Dere} {et~al.}(1997){Dere}, {Landi}, {Mason}, {Monsignori Fossi}, \&
  {Young}}]{1997A&AS..125..149D}
{Dere}, K.~P., {Landi}, E., {Mason}, H.~E., {Monsignori Fossi}, B.~C., \&
  {Young}, P.~R. 1997, \aaps, 125, 149

\bibitem[{{Fontaine} {et~al.}(2001){Fontaine}, {Brassard}, \&
  {Bergeron}}]{2001PASP..113..409F}
{Fontaine}, G., {Brassard}, P., \& {Bergeron}, P. 2001, \pasp, 113, 409

\bibitem[{{Foreman-Mackey}(2016)}]{Foreman-Mackey2016}
{Foreman-Mackey}, D. 2016, The Journal of Open Source Software, 1, 24

\bibitem[{{Foreman-Mackey}(2017)}]{corner_ascl}
{Foreman-Mackey}, D. 2017, {corner.py: Corner plots}, Astrophysics Source Code
  Library, record ascl:1702.002

\bibitem[{{Fruscione} {et~al.}(2006){Fruscione}, {McDowell}, {Allen},
  {Brickhouse}, {Burke}, {Davis}, {Durham}, {Elvis}, {Galle}, {Harris},
  {Huenemoerder}, {Houck}, {Ishibashi}, {Karovska}, {Nicastro}, {Noble},
  {Nowak}, {Primini}, {Siemiginowska}, {Smith}, \& {Wise}}]{Fruscione_etal2006}
{Fruscione}, A., {McDowell}, J.~C., {Allen}, G.~E., {et~al.} 2006, in Society
  of Photo-Optical Instrumentation Engineers (SPIE) Conference Series, Vol.
  6270, Society of Photo-Optical Instrumentation Engineers (SPIE) Conference
  Series, ed. D.~R. {Silva} \& R.~E. {Doxsey}, 62701V

\bibitem[{{G{\"a}nsicke} {et~al.}(1998){G{\"a}nsicke}, {van Teeseling},
  {Beuermann}, \& {de Martino}}]{1998A&A...333..163G}
{G{\"a}nsicke}, B.~T., {van Teeseling}, A., {Beuermann}, K., \& {de Martino},
  D. 1998, \aap, 333, 163

\bibitem[{{Greiner} \& {Di Stefano}(2002)}]{2002A&A...387..944G}
{Greiner}, J. \& {Di Stefano}, R. 2002, \aap, 387, 944

\bibitem[{{Greiner} {et~al.}(1991){Greiner}, {Hasinger}, \&
  {Kahabka}}]{1991A&A...246L..17G}
{Greiner}, J., {Hasinger}, G., \& {Kahabka}, P. 1991, \aap, 246, L17

\bibitem[{{Griem}(1960)}]{1960ApJ...132..883G}
{Griem}, H.~R. 1960, \apj, 132, 883

\bibitem[{{Griem}(1967)}]{1967ApJ...148..547G}
{Griem}, H.~R. 1967, \apj, 148, 547

\bibitem[{{Hartmann} \& {Heise}(1997)}]{1997A&A...322..591H}
{Hartmann}, H.~W. \& {Heise}, J. 1997, \aap, 322, 591

\bibitem[{{Hartmann} {et~al.}(1999){Hartmann}, {Heise}, {Kahabka}, {Motch}, \&
  {Parmar}}]{1999A&A...346..125H}
{Hartmann}, H.~W., {Heise}, J., {Kahabka}, P., {Motch}, C., \& {Parmar}, A.~N.
  1999, \aap, 346, 125

\bibitem[{{Hauschildt} {et~al.}(1997){Hauschildt}, {Baron}, \&
  {Allard}}]{1997ApJ...483..390H}
{Hauschildt}, P.~H., {Baron}, E., \& {Allard}, F. 1997, \apj, 483, 390

\bibitem[{{Heise} {et~al.}(1994){Heise}, {van Teeseling}, \&
  {Kahabka}}]{1994A&A...288L..45H}
{Heise}, J., {van Teeseling}, A., \& {Kahabka}, P. 1994, \aap, 288, L45

\bibitem[{{Hubeny}(1988)}]{1988CoPhC..52..103H}
{Hubeny}, I. 1988, Computer Physics Communications, 52, 103

\bibitem[{{Hubeny} {et~al.}(1994){Hubeny}, {Hummer}, \&
  {Lanz}}]{1994A&A...282..151H}
{Hubeny}, I., {Hummer}, D.~G., \& {Lanz}, T. 1994, \aap, 282, 151

\bibitem[{{Huenemoerder} {et~al.}(2011){Huenemoerder}, {Mitschang}, {Dewey},
  {Nowak}, {Schulz}, {Nichols}, {Davis}, {Houck}, {Marshall}, {Noble},
  {Morgan}, \& {Canizares}}]{Huenemoerder_etal2011}
{Huenemoerder}, D.~P., {Mitschang}, A., {Dewey}, D., {et~al.} 2011, \aj, 141,
  129

\bibitem[{{Hummer} \& {Mihalas}(1988)}]{1988ApJ...331..794H}
{Hummer}, D.~G. \& {Mihalas}, D. 1988, \apj, 331, 794

\bibitem[{{Iben}(1982)}]{1982ApJ...259..244I}
{Iben}, I., J. 1982, \apj, 259, 244

\bibitem[{{Ibragimov} {et~al.}(2003){Ibragimov}, {Suleimanov}, {Vikhlinin}, \&
  {Sakhibullin}}]{2003ARep...47..186I}
{Ibragimov}, A.~A., {Suleimanov}, V.~F., {Vikhlinin}, A., \& {Sakhibullin},
  N.~A. 2003, Astronomy Reports, 47, 186

\bibitem[{{Kahabka}(1995)}]{1995A&A...304..227K}
{Kahabka}, P. 1995, \aap, 304, 227

\bibitem[{{Kahabka}(1997)}]{1997ASPC..121..730K}
{Kahabka}, P. 1997, in Astronomical Society of the Pacific Conference Series,
  Vol. 121, IAU Colloq. 163: Accretion Phenomena and Related Outflows, ed.
  D.~T. {Wickramasinghe}, G.~V. {Bicknell}, \& L.~{Ferrario}, 730

\bibitem[{{Kahabka}(1998)}]{1998A&A...331..328K}
{Kahabka}, P. 1998, \aap, 331, 328

\bibitem[{{Kahabka} {et~al.}(1999){Kahabka}, {Hartmann}, {Parmar}, \&
  {Negueruela}}]{1999A&A...347L..43K}
{Kahabka}, P., {Hartmann}, H.~W., {Parmar}, A.~N., \& {Negueruela}, I. 1999,
  \aap, 347, L43

\bibitem[{{Kahabka} \& {van den Heuvel}(1997)}]{1997ARA&A..35...69K}
{Kahabka}, P. \& {van den Heuvel}, E.~P.~J. 1997, \araa, 35, 69

\bibitem[{{Karzas} \& {Latter}(1961)}]{1961ApJS....6..167K}
{Karzas}, W.~J. \& {Latter}, R. 1961, \apjs, 6, 167

\bibitem[{{K{\"o}nig} {et~al.}(2022){K{\"o}nig}, {Wilms}, {Arcodia}, {Dauser},
  {Dennerl}, {Doroshenko}, {Haberl}, {H{\"a}mmerich}, {Kirsch}, {Kreykenbohm},
  {Lorenz}, {Malyali}, {Merloni}, {Rau}, {Rauch}, {Sala}, {Schwope},
  {Suleimanov}, {Weber}, \& {Werner}}]{2022Natur.605..248K}
{K{\"o}nig}, O., {Wilms}, J., {Arcodia}, R., {et~al.} 2022, \nat, 605, 248

\bibitem[{{Kurucz}(1993{\natexlab{a}})}]{1993KurCD..13.....K}
{Kurucz}, R. 1993{\natexlab{a}}, ATLAS9 Stellar Atmosphere Programs and 2 km/s
  grid. Kurucz CD-ROM No. 13. Cambridge, 13

\bibitem[{{Kurucz}(1970)}]{1970SAOSR.309.....K}
{Kurucz}, R.~L. 1970, SAO Special Report, 309

\bibitem[{{Kurucz}(1993{\natexlab{b}})}]{1993yCat.6039....0K}
{Kurucz}, R.~L. 1993{\natexlab{b}}, VizieR Online Data Catalog, VI/39

\bibitem[{{Lanz} {et~al.}(2005){Lanz}, {Telis}, {Audard}, {Paerels},
  {Rasmussen}, \& {Hubeny}}]{2005ApJ...619..517L}
{Lanz}, T., {Telis}, G.~A., {Audard}, M., {et~al.} 2005, \apj, 619, 517

\bibitem[{{Long} {et~al.}(1981){Long}, {Helfand}, \&
  {Grabelsky}}]{1981ApJ...248..925L}
{Long}, K.~S., {Helfand}, D.~J., \& {Grabelsky}, D.~A. 1981, \apj, 248, 925

\bibitem[{{McGowan} {et~al.}(2005){McGowan}, {Charles}, {Blustin}, {Livio},
  {O'Donoghue}, \& {Heathcote}}]{2005MNRAS.364..462M}
{McGowan}, K.~E., {Charles}, P.~A., {Blustin}, A.~J., {et~al.} 2005, \mnras,
  364, 462

\bibitem[{{Mihalas}(1978)}]{1978stat.book.....M}
{Mihalas}, D. 1978, {Stellar atmospheres} (San Francisco: W.H. Freeman)

\bibitem[{{Nauenberg}(1972)}]{1972ApJ...175..417N}
{Nauenberg}, M. 1972, \apj, 175, 417

\bibitem[{{Ness}(2010)}]{2010AN....331..179N}
{Ness}, J.~U. 2010, Astronomische Nachrichten, 331, 179

\bibitem[{{Ness}(2020)}]{2020AdSpR..66.1202N}
{Ness}, J.-U. 2020, Advances in Space Research, 66, 1202

\bibitem[{{Ness} {et~al.}(2013){Ness}, {Osborne}, {Henze}, {Dobrotka}, {Drake},
  {Ribeiro}, {Starrfield}, {Kuulkers}, {Behar}, {Hernanz}, {Schwarz}, {Page},
  {Beardmore}, \& {Bode}}]{2013A&A...559A..50N}
{Ness}, J.~U., {Osborne}, J.~P., {Henze}, M., {et~al.} 2013, \aap, 559, A50

\bibitem[{{Ness} {et~al.}(2003){Ness}, {Starrfield}, {Burwitz}, {Wichmann},
  {Hauschildt}, {Drake}, {Wagner}, {Bond}, {Krautter}, {Orio}, {Hernanz},
  {Gehrz}, {Woodward}, {Butt}, {Mukai}, {Balman}, \&
  {Truran}}]{2003ApJ...594L.127N}
{Ness}, J.~U., {Starrfield}, S., {Burwitz}, V., {et~al.} 2003, \apjl, 594, L127

\bibitem[{{Nomoto} {et~al.}(2007){Nomoto}, {Saio}, {Kato}, \&
  {Hachisu}}]{2007ApJ...663.1269N}
{Nomoto}, K., {Saio}, H., {Kato}, M., \& {Hachisu}, I. 2007, \apj, 663, 1269

\bibitem[{{Odendaal} {et~al.}(2014){Odendaal}, {Meintjes}, {Charles}, \&
  {Rajoelimanana}}]{2014MNRAS.437.2948O}
{Odendaal}, A., {Meintjes}, P.~J., {Charles}, P.~A., \& {Rajoelimanana}, A.~F.
  2014, \mnras, 437, 2948

\bibitem[{{Orio} {et~al.}(2001){Orio}, {Covington}, \&
  {{\"O}gelman}}]{2001A&A...373..542O}
{Orio}, M., {Covington}, J., \& {{\"O}gelman}, H. 2001, \aap, 373, 542

\bibitem[{{Orio} {et~al.}(2022){Orio}, {Gendreau}, {Giese}, {Luna}, {Magdolen},
  {Pei}, {Sun}, {Behar}, {Dobrotka}, {Mikolajewska}, {Pasham}, \&
  {Strohmayer}}]{2022ApJ...932...45O}
{Orio}, M., {Gendreau}, K., {Giese}, M., {et~al.} 2022, \apj, 932, 45

\bibitem[{{Paerels} {et~al.}(2001){Paerels}, {Rasmussen}, {Hartmann}, {Heise},
  {Brinkman}, {de Vries}, \& {den Herder}}]{2001A&A...365L.308P}
{Paerels}, F., {Rasmussen}, A.~P., {Hartmann}, H.~W., {et~al.} 2001, \aap, 365,
  L308

\bibitem[{{Pakull} {et~al.}(1993){Pakull}, {Motch}, {Bianchi}, {Thomas},
  {Guibert}, {Beaulieu}, {Grison}, \& {Schaeidt}}]{1993A&A...278L..39P}
{Pakull}, M.~W., {Motch}, C., {Bianchi}, L., {et~al.} 1993, \aap, 278, L39

\bibitem[{{Parmar} {et~al.}(1998){Parmar}, {Kahabka}, {Hartmann}, {Heise}, \&
  {Taylor}}]{1998A&A...332..199P}
{Parmar}, A.~N., {Kahabka}, P., {Hartmann}, H.~W., {Heise}, J., \& {Taylor},
  B.~G. 1998, \aap, 332, 199

\bibitem[{{Pietrzy{\'n}ski} {et~al.}(2019){Pietrzy{\'n}ski}, {Graczyk},
  {Gallenne}, {Gieren}, {Thompson}, {Pilecki}, {Karczmarek}, {G{\'o}rski},
  {Suchomska}, {Taormina}, {Zgirski}, {Wielg{\'o}rski}, {Ko{\l}aczkowski},
  {Konorski}, {Villanova}, {Nardetto}, {Kervella}, {Bresolin}, {Kudritzki},
  {Storm}, {Smolec}, \& {Narloch}}]{2019Natur.567..200P}
{Pietrzy{\'n}ski}, G., {Graczyk}, D., {Gallenne}, A., {et~al.} 2019, \nat, 567,
  200

\bibitem[{{Rauch}(2003)}]{2003A&A...403..709R}
{Rauch}, T. 2003, \aap, 403, 709

\bibitem[{{Rauch} {et~al.}(2010){Rauch}, {Orio}, {Gonzales-Riestra}, {Nelson},
  {Still}, {Werner}, \& {Wilms}}]{2010ApJ...717..363R}
{Rauch}, T., {Orio}, M., {Gonzales-Riestra}, R., {et~al.} 2010, \apj, 717, 363

\bibitem[{{Rauch} \& {Werner}(2010)}]{2010AN....331..146R}
{Rauch}, T. \& {Werner}, K. 2010, Astronomische Nachrichten, 331, 146

\bibitem[{{Rolleston} {et~al.}(2002){Rolleston}, {Trundle}, \&
  {Dufton}}]{2002A&A...396...53R}
{Rolleston}, W.~R.~J., {Trundle}, C., \& {Dufton}, P.~L. 2002, \aap, 396, 53

\bibitem[{{Schaeidt} {et~al.}(1993){Schaeidt}, {Hasinger}, \&
  {Truemper}}]{1993A&A...270L...9S}
{Schaeidt}, S., {Hasinger}, G., \& {Truemper}, J. 1993, \aap, 270, L9

\bibitem[{{Schwarz} {et~al.}(2011){Schwarz}, {Ness}, {Osborne}, {Page},
  {Evans}, {Beardmore}, {Walter}, {Helton}, {Woodward}, {Bode}, {Starrfield},
  \& {Drake}}]{2011ApJS..197...31S}
{Schwarz}, G.~J., {Ness}, J.-U., {Osborne}, J.~P., {et~al.} 2011, \apjs, 197,
  31

\bibitem[{{Seaton} {et~al.}(1994){Seaton}, {Yan}, {Mihalas}, \&
  {Pradhan}}]{1994MNRAS.266..805S}
{Seaton}, M.~J., {Yan}, Y., {Mihalas}, D., \& {Pradhan}, A.~K. 1994, \mnras,
  266, 805

\bibitem[{{Skopal}(2022)}]{2022AJ....164..145S}
{Skopal}, A. 2022, \aj, 164, 145

\bibitem[{{Southwell} {et~al.}(1996){Southwell}, {Livio}, {Charles},
  {O'Donoghue}, \& {Sutherland}}]{1996ApJ...470.1065S}
{Southwell}, K.~A., {Livio}, M., {Charles}, P.~A., {O'Donoghue}, D., \&
  {Sutherland}, W.~J. 1996, \apj, 470, 1065

\bibitem[{{Stecchini} {et~al.}(2023){Stecchini}, {Perez Diaz}, {D'Amico}, \&
  {Jablonski}}]{2023MNRAS.522.3472S}
{Stecchini}, P.~E., {Perez Diaz}, M., {D'Amico}, F., \& {Jablonski}, F. 2023,
  \mnras, 522, 3472

\bibitem[{{Suleimanov} {et~al.}(2014{\natexlab{a}}){Suleimanov}, {Hertfelder},
  {Werner}, \& {Kley}}]{2014A&A...571A..55S}
{Suleimanov}, V., {Hertfelder}, M., {Werner}, K., \& {Kley}, W.
  2014{\natexlab{a}}, \aap, 571, A55

\bibitem[{{Suleimanov} \& {Ibragimov}(2003)}]{2003ARep...47..197S}
{Suleimanov}, V.~F. \& {Ibragimov}, A.~A. 2003, Astronomy Reports, 47, 197

\bibitem[{{Suleimanov} {et~al.}(2014{\natexlab{b}}){Suleimanov}, {Klochkov},
  {Pavlov}, \& {Werner}}]{2014ApJS..210...13S}
{Suleimanov}, V.~F., {Klochkov}, D., {Pavlov}, G.~G., \& {Werner}, K.
  2014{\natexlab{b}}, \apjs, 210, 13

\bibitem[{{Suleimanov} {et~al.}(2013){Suleimanov}, {Mauche}, {Zhuchkov}, \&
  {Werner}}]{2013ASPC..469..349S}
{Suleimanov}, V.~F., {Mauche}, C.~W., {Zhuchkov}, R.~Y., \& {Werner}, K. 2013,
  in Astronomical Society of the Pacific Conference Series, Vol. 469, 18th
  European White Dwarf Workshop., ed. J.~{Krzesi{\'n}ski}, G.~{Stachowski},
  P.~{Moskalik}, \& K.~{Bajan}, 349

\bibitem[{{Sutherland}(1998)}]{1998MNRAS.300..321S}
{Sutherland}, R.~S. 1998, \mnras, 300, 321

\bibitem[{{Swartz} {et~al.}(2002){Swartz}, {Ghosh}, {Suleimanov}, {Tennant}, \&
  {Wu}}]{2002ApJ...574..382S}
{Swartz}, D.~A., {Ghosh}, K.~K., {Suleimanov}, V., {Tennant}, A.~F., \& {Wu},
  K. 2002, \apj, 574, 382

\bibitem[{{Tr{\"u}mper} {et~al.}(1991){Tr{\"u}mper}, {Hasinger}, {Aschenbach},
  {Br{\"a}uninger}, {Briel}, {Burkert}, {Fink}, {Pfeffermann}, {Pietsch},
  {Predehl}, {Schmitt}, {Voges}, {Zimmermann}, \&
  {Beuermann}}]{1991Natur.349..579T}
{Tr{\"u}mper}, J., {Hasinger}, G., {Aschenbach}, B., {et~al.} 1991, \nat, 349,
  579

\bibitem[{{van den Heuvel} {et~al.}(1992){van den Heuvel}, {Bhattacharya},
  {Nomoto}, \& {Rappaport}}]{1992A&A...262...97V}
{van den Heuvel}, E.~P.~J., {Bhattacharya}, D., {Nomoto}, K., \& {Rappaport},
  S.~A. 1992, \aap, 262, 97

\bibitem[{{van Rossum}(2012)}]{2012ApJ...756...43V}
{van Rossum}, D.~R. 2012, \apj, 756, 43

\bibitem[{{van Rossum} \& {Ness}(2010)}]{2010AN....331..175V}
{van Rossum}, D.~R. \& {Ness}, J.~U. 2010, Astronomische Nachrichten, 331, 175

\bibitem[{{Verner} {et~al.}(1996){Verner}, {Ferland}, {Korista}, \&
  {Yakovlev}}]{1996ApJ...465..487V}
{Verner}, D.~A., {Ferland}, G.~J., {Korista}, K.~T., \& {Yakovlev}, D.~G. 1996,
  \apj, 465, 487

\bibitem[{{Werner} {et~al.}(2003){Werner}, {Deetjen}, {Dreizler}, {Nagel},
  {Rauch}, \& {Schuh}}]{2003ASPC..288...31W}
{Werner}, K., {Deetjen}, J.~L., {Dreizler}, S., {et~al.} 2003, in Astronomical
  Society of the Pacific Conference Series, Vol. 288, Stellar Atmosphere
  Modeling, ed. I.~{Hubeny}, D.~{Mihalas}, \& K.~{Werner}, 31

\bibitem[{{Wolf} {et~al.}(2013){Wolf}, {Bildsten}, {Brooks}, \&
  {Paxton}}]{2013ApJ...777..136W}
{Wolf}, W.~M., {Bildsten}, L., {Brooks}, J., \& {Paxton}, B. 2013, \apj, 777,
  136

\end{thebibliography}

\end{document}